\documentclass[traditabstract]{aa}
\usepackage{graphicx}
\usepackage{txfonts}
\usepackage{natbib}
\usepackage{color}
\bibpunct{(}{)}{;}{a}{}{,}

\newcommand{\Sr}[5]{\mbox{${\rm #1}\,^#2{\rm #3}^{{\rm #4}}_{\rm #5}$}}

\newcommand{\Teff}{T_{\rm eff}}
\newcommand{\EW}{W_{\lambda}}
\newcommand{\mA}{{\rm m\AA}}
\newcommand{\Elow}{E_{\rm low}}
\newcommand{\Eup}{E_{\rm up}}
\newcommand{\Vmic}{\xi_{\rm t}}

\newcommand{\opd}{\log \tau_{\rm c}}

\newcommand{\drs}{\Delta_{\rm NLTE}}
\begin{document}
\title{NLTE analysis of Sr lines in spectra of late-type stars with new R-matrix
atomic data}
\author{M. Bergemann\inst{1}, C. J. Hansen\inst{2}, M. Bautista\inst{3}, and G.
Ruchti\inst{1}}
\institute{Max-Planck Institute for Astrophysics, Karl-Schwarzschild Str. 1,
85741, Garching, Germany  \and Landessternwarte, K\"onigstuhl 12, 69117
Heidelberg, Germany \and Department of Physics, Western Michigan University,
Kalamazoo, MI 49008, USA}
\date{Received date / Accepted date}
\abstract{
We investigate statistical equilibrium of neutral and singly-ionized strontium
in late-type stellar atmospheres. Particular attention is given to the
completeness of the model atom, which includes new energy levels, transition
probabilities, photoionization and electron-impact excitation cross-sections
computed with the R-matrix method. The NLTE model is applied to the analysis of
Sr I and Sr II lines in the spectra of the Sun, Procyon, Arcturus, and HD
122563, showing a significant improvement in the ionization balance compared to
LTE line formation calculations, which predict abundance discrepancies of up to
$0.5$ dex. The solar Sr abundance is $\log \epsilon = 2.93 \pm 0.04$ dex, in
agreement with the meteorites. A grid of NLTE abundance corrections for Sr I and
Sr II lines covering a large range of stellar parameters is presented.}
\titlerunning{Statistical equilibrium of Sr in late-type stars}
\authorrunning{M. Bergemann et al.}
\maketitle
%
%
\section{Introduction}
Spectroscopic observations of low-mass stars have shaped our understanding
of Galactic evolution and stellar nucleosynthesis. Strontium, as one of the
abundant r-process elements, has been extensively investigated in the past few
decades. However, its main production site has not yet been identified: the
observed abundances of Sr in metal-poor stars are far too large to be
explained by conventional rapid neutron-capture nucleosynthesis in SNe II,
suggesting some alternative exotic scenarios, such as the light element primary
process \citep{2004A&A...425.1029T}, rp-process in accretion disks around
low-mass black holes \citep{1998PhR...294..167S}, black hole - neutron star
mergers \citep{2008ApJ...679L.117S}, high-entropy winds in SN II 
\citep{2010ApJ...712.1359F}, and low-mass electron-capture supernovae
\citep{2011ApJ...726L..15W} to name just a few \citep[see][and references
therein]{2011RPPh...74i6901J}.

Until recently, determinations of Sr abundances in metal-poor stars relied
almost exclusively on the two near-UV lines of \ion{Sr}{ii}, which are
sufficiently strong to be detected also in the spectra of moderate-to-low
resolution and signal-to-noise. The drawback is that in spectra of stars
typically used for studies of Galactic chemical evolution, [Fe/H] $ > -2$, these
lines saturate and develop pronounced damping wings, overlapping with various
atomic and molecular blends. Thus, at higher metallicities a preference is
sometimes given to the weak \ion{Sr}{i} line at $4607.34$ \AA\ and/or the 
\ion{Sr}{ii} line at $4161$ \AA. However, there is evidence that the
\ion{Sr}{i} lines may be subject to non-local thermodynamic equilibrium
(hereafter, NLTE) effects \citep{2000MNRAS.311..535B}, which has been supported
by \textit{ab initio} calculations solving for radiative transfer in NLTE for a
small grid or red giant model atmospheres \citep{2006ApJ...641..494S}. In a few
studies utilizing near-IR spectra, also the \ion{Sr}{ii} triplet
($10\,037$, $10\,327$, and $10\,915$ $\AA$ have been used
\citep{2011A&A...530A.105A}. In the majority of published studies, the
preference is given to one ionization stage only
\citep[e.g.,][]{1999A&A...341..241J}, and only a few studies
investigated both \ion{Sr}{i} and \ion{Sr}{ii} lines
\citep{1994A&A...287..927G, 2002ApJ...572..861C}, finding discrepant results. 

In this study, we perform for the first time a NLTE analysis of the \ion{Sr}{i}
and \ion{Sr}{ii} lines in spectra of late-type stars. The new atomic model was
constructed from the state-of-art atomic data, computed specifically for this
work. The NLTE model atom is tested on a number of reference stars with
parameters determined by other independent methods. Furthermore, we present a
large grid of NLTE abundance corrections for \ion{Sr}{i} and \ion{Sr}{ii} lines.
The results presented in this work will be applied to the analysis of a
representative sample of metal-poor stars observed at a very-high resolution and
signal-to-noise in Hansen et al. (in prep.) Furthermore, we plan to undertake the
NLTE Sr abundance analysis of the thick-disk and halo stars with
spectroscopic parameters from \citet{2011ApJ...737....9R}.
\section{Methods}{\label{sec:methods}}
\subsection{Model atmospheres and NLTE calculations}{\label{sec:nlte}}
All calculations in this work were performed with classical 1D LTE
plane-parallel model atmospheres. We used the MAFAGS-OS
\citep{2004A&A...420..289G,2004A&A...426..309G} and MAFAGS-ODF models, which are
well-adapted for the analysis of late-type stars. A brief description of these
models and comparison with other models of a similar type
\citep[MARCS,][]{2008A&A...486..951G} is presented in \citet{bergemann12}.

The NLTE statistical equilibrium calculations were performed with the revised
version of the DETAIL code \citep{butler85}. The statistical equilibrium and
radiative transfer equations are solved by the accelerated $\Lambda$-iteration
method in the formulation of \citet{1991A&A...245..171R,1992A&A...262..209R}.
The method allows for self-consistent treatment of overlapping lines and
continua. A description of the code with some recent modifications related to
the treatment of background line opacity can be found in \citet{bergemann12}. In
the statistical equilibrium calculations, we treat the diagnostic lines 
investigated here with Voigt profiles, whereas all other Sr lines were computed
with a Gaussian profile with $13$ frequency points.

The LTE and NLTE abundances for the reference stars, as well as the NLTE
abundance corrections, were determined by full spectrum synthesis with the
revised version of the SIU code \citep{reetz}, which was adapted for automated
NLTE abundance calculations (L. Sbordone, private communication). The
line lists in SIU have been continuously updated by the members of the Munich
group \citep{2004A&A...413.1045G, 2006A&A...451.1065G} and they are mainly based
on the Kurucz\footnote{kurucz.harvard.edu/}, Hannover \footnote{www.pmp.uni-hannover.de/cgi-bin/ssi/test/kurucz/sekur.html. The atomic line data used in this database are taken from \citet{1995KurCD23K}.}, and NIST \citep{nist} compilations.
\subsection{Model atom}{\label{sec:modelatom}}
The model atom of Sr was constructed using the available atomic data
from the Hannover and NIST databases and supplemented by our new calculations\footnote{The new atomic data for Sr presented in the paper
and in Bautista et al. (2002), including the energy levels, $f$-values,
photoionization and e-impact excitation cross-sections, can be provided upon
request.} of atomic energy levels, dipole-permitted and forbidden transitions
(see below). Thus, the model is the most complete representation of atomic
system of \ion{Sr}{i} and \ion{Sr}{ii}, to date. In particular, we also include
$135$ \ion{Sr}{ii} transitions from the Coulomb approximation
calculations of \citet{1977ADNDT..19..533L}. The neutral atom is, thus,
represented by $141$ levels with the principle quantum numbers n$* \leq 20$, and
the uppermost level $20$f \element[][1]{F^\circ} is located at $5.66$ eV, i.e.
$0.03$ eV from the 1-st ionization limit at $5.69$ eV. The singly-ionized atom
includes $49$ levels including the highest experimentally-observed
level $14$d \element[][2]{D}, which is located at $10.684$ eV. The model is closed by the \ion{Sr}{iii} ground state 4p$^6$ \element[][1]{S_0}. The number of
dipole-allowed transitions is $336$ and $214$ for \ion{Sr}{i} and
\ion{Sr}{ii}, respectively.
\begin{figure}
\centering
\resizebox{\columnwidth}{!}{\rotatebox{90}{\includegraphics{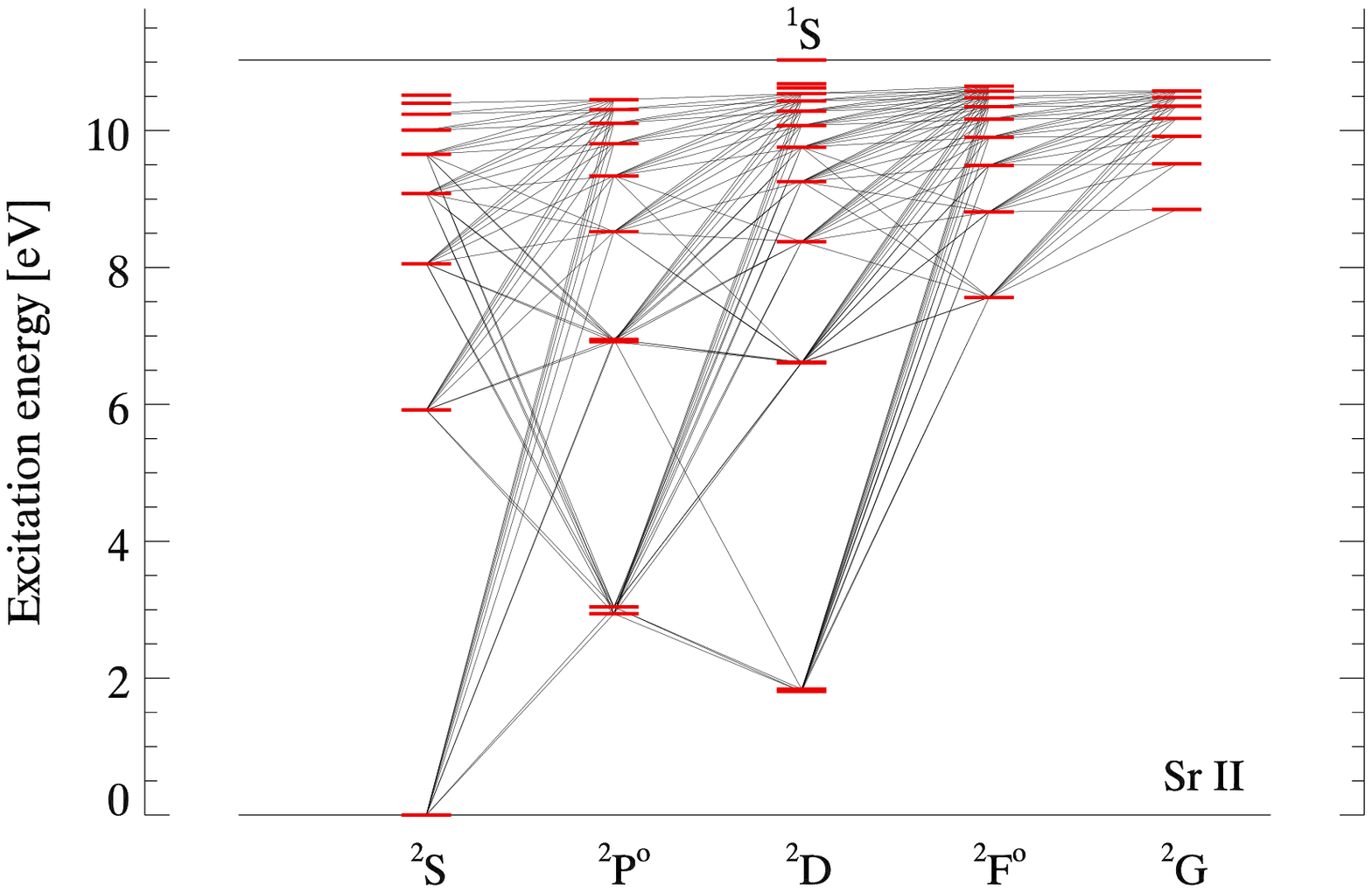}}}
\resizebox{\columnwidth}{!}{\rotatebox{90}{\includegraphics{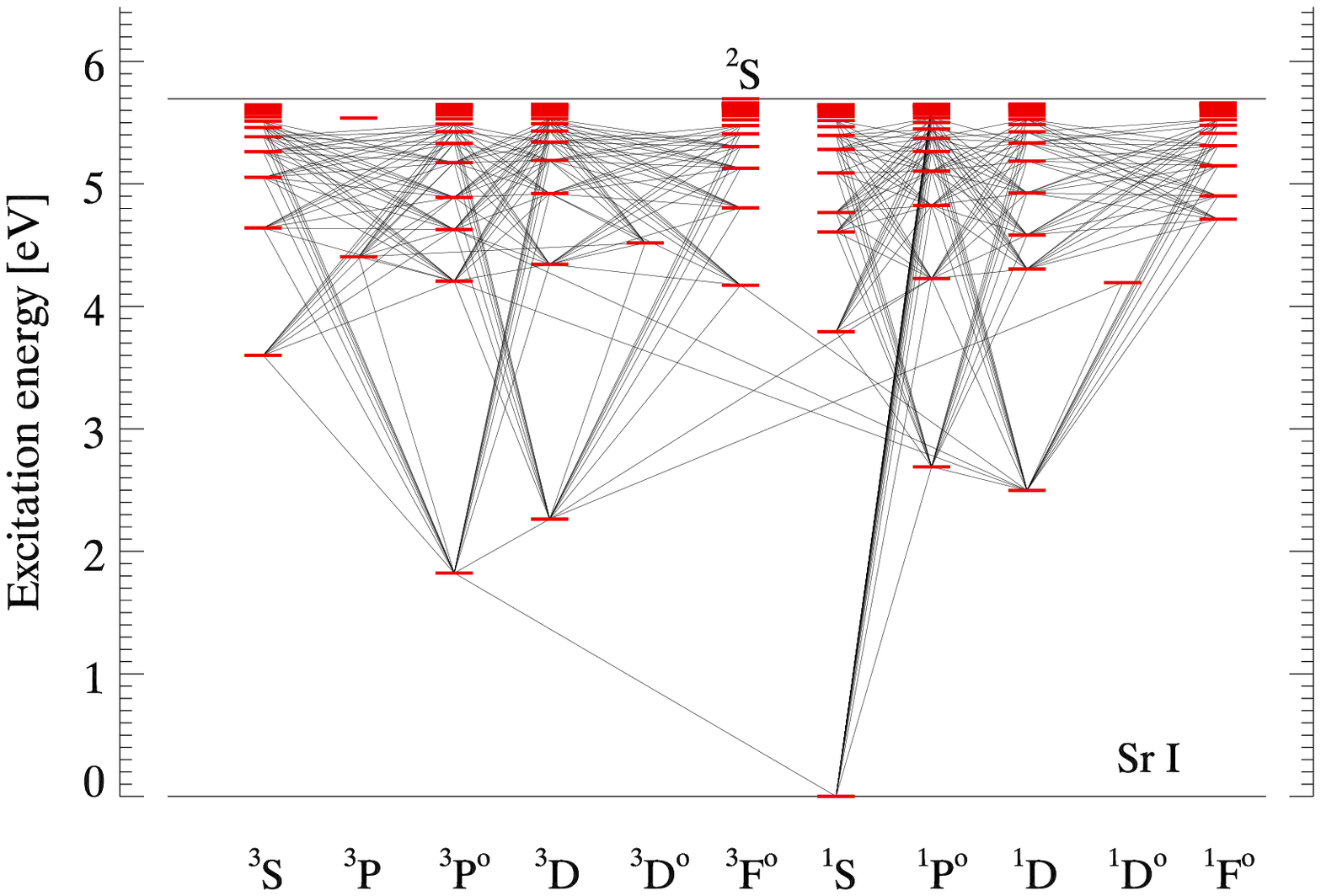}}}
\caption{Grotrian diagram of the \ion{Sr}{i} (bottom) and \ion{Sr}{ii} (top)
model atoms. Only dipole-allowed transitions are shown. This is the most
complete model atom utilized for NLTE calculations of FGK stars so far.}
\label{atom}
\end{figure}

Our model of Sr is similar to that of \citet{2011A&A...530A.105A} with
respect to the term structure and the \textit{number} of dipole-permitted
transitions of \ion{Sr}{ii}. However, we include more accurate transition
probabilities for \ion{Sr}{ii}. More importantly, our model atom includes new
quantum-mechanical photoionization cross-sections for \ion{Sr}{i} and e$^-$
impact excitation cross-sections for dipole-permitted and forbidden transitions
in \ion{Sr}{ii} (see below). The model atoms employed by
\citet{2006ApJ...641..494S} and \citet{2001A&A...376..232M} are much simpler
than that employed in this work ($84$ and $41$ total levels, respectively).

\subsubsection{New calculations of atomic data}

We compute single-electron orbitals for the target ions\footnote{We note that
different notations are standard for atomic physics and spectroscopy, and we
chose to respect this difference in our work. Ionization stage of an element is
thus indicated by a numerical superscript in Sect. 2.2.1 describing the atomic
physics calculations and by a roman numerical in astrophysical application in
the further sections.}  Sr$^+$ with the atomic structure code
\textit{autostructure} by \citet{bad86,bad97}. This code is an extension of the
program superstructure \citep{1969JPhB....2.1028E}, computes
fine-structure CI energy levels, and radiative and Auger rates in a Breit-Pauli
relativistic framework. Single electron orbitals are constructed by
diagonalizing the non-relativistic Hamiltonian, within a statistical
Thomas-Fermi-Dirac-Amaldi (TFDA) model potential \citep{1969JPhB....2.1028E}.
The scaling parameters are optimized by minimizing a weighted sum of the LS term
energies. We employ a very extensive configuration expansion with
configurations of the form $4p^6nl$, with $n$ going from 4 to 6 and $l \le 3$,
and configurations with multiple promotion from the 4p orbital, i.e.
$4p^q5s^r4d^tnl$ with $3\le q \le 6$, $0 \le r \le 2$, and $0 \le t \le 2$.
These multiple promotions from the 4p orbital are essential in computing the
photoionization cross sections and that will be the subject of a future
publication.

Our target representations give term energies in reasonable agreement, within
$\sim5$ percent with experimental values.

\begin{figure}
\centering
\resizebox{\columnwidth}{!}{{\includegraphics{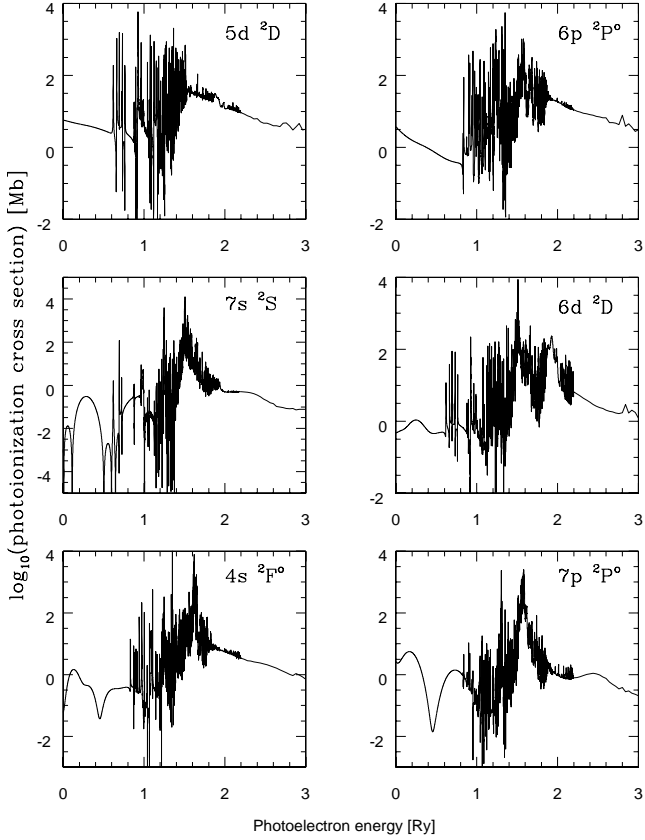}}}
\caption{Partial photoionization cross-sections for the \ion{Sr}{i} ground state
into the lowest six states of the \ion{Sr}{ii} target. The cross-sections are
dominated by prominent resonances.}
\label{photo_par}
\end{figure}

\begin{figure}
\centering
\resizebox{\columnwidth}{!}{{\includegraphics{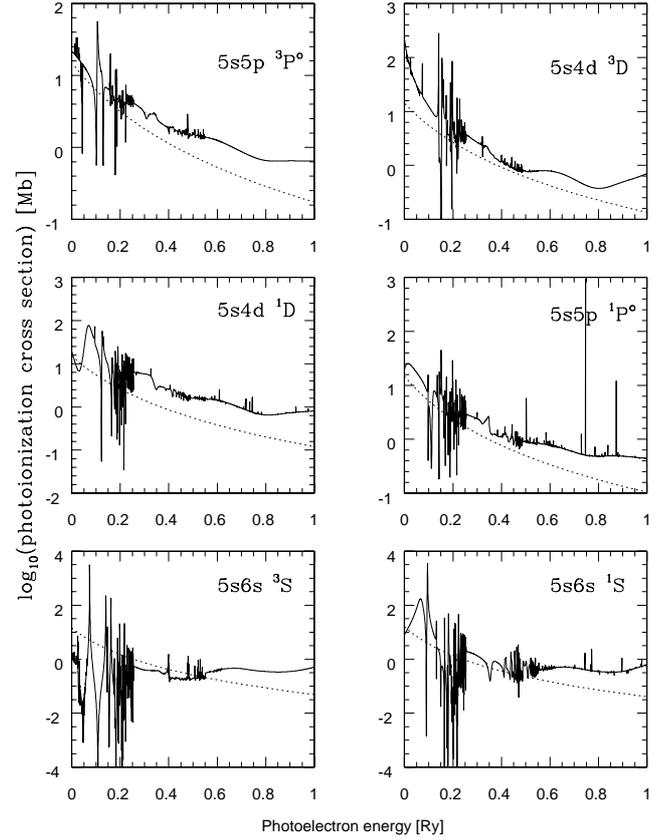}}}
\caption{Total photoionization cross-sections for the selected \ion{Sr}{i}
levels. Dotted lines show hydrogenic cross-sections. The new
cross-sections are up to few orders of magnitude larger than the commonly-used
hydrogenic approximation.}
\label{photo_tot}
\end{figure}

The photoionization cross-sections are computed with the R-matrix method
\citep{burke71}. The present calculation for Sr$^0$ includes the lowest $67$
LS-terms of the target in the close coupling expansion and all short range
(N+1)-electron configurations that results from adding an electron to the target
configurations. We find $197$ singlet and triplet bound terms of Sr$^0$ with $n
< 9$ and $L \leq 5$, and compute photoionization cross sections for all of them.
The cross sections are computed at an even energy mesh of $5 \times 10^{-5}$ Ry
from threshold up to 0.23 Ry, mesh of $10^{-3}$ Ry from 0.23 Ry to 2.3 Ry
(roughly the highest target threshold), and another mesh of 0.02 Ry from 2.3 Ry
to 4.3 Ry. Figure \ref{photo_par} shows the partial cross sections for
ionization from the ground term of Sr$^0$ into each of the first six terms of
the target Sr$^+$. We note, however, that DETAIL is not yet able to consistently
treat ionization to specific states of the target ion. Therefore, total
photoionization cross-sections are adopted here. Figure \ref{photo_tot} shows
the total cross sections for the lowest six excited state of Sr$^0$. For
comparison, we show the hydrogenic cross-sections computed with effective
principal quantum number. The latter reproduce the functional dependence of the
cross-section with energy. However, the quantum-mechanical photoionization
cross-sections are typically larger both in the background (factor of two to
five) and in the resonances (up to three orders of magnitude).
\subsubsection{Collision rates}
Electron impact excitation rate coefficients for the $49$ levels of \ion{Sr}{ii}
levels were taken from \citet{2002MNRAS.331..875B}. The data were computed with
the same technique, i.e., the close-coupling approximation with the R-matrix
method. For all other levels of \ion{Sr}{i} and \ion{Sr}{ii} the electron
collision rates were computed using the formulae of \citet{1962ApJ...136..906V}
for dipole-allowed and \citet{1973asqu.book.....A} for
dipole-forbidden transitions. Cross-sections for transitions caused by inelastic
collisions with \ion{H}{i} atoms are basically unknown for any atom heavier than
Mg. The only available formula developed originally for collisions between equal
H-like atoms \citep{1968ZPhy..211..404D, 1969ZPhy..225..470D} and later slightly
modified by \citet{1984A&A...130..319S} and \citet{1993PhST...47..186L} was
shown to over-estimate the rates of bound-bound transitions by two to seven
orders of magnitude \citep{2003PhRvA..68f2703B, barklem12}. Furthermore, the
charge transfer processes can not be described by this simple classical formula
at all. Our tests with various scaling factors to the Drawin's \ion{H}{i}
inelastic collision cross-sections demonstrated that excitation and ionization
balance of \ion{Sr}{i}/\ion{Sr}{ii} in the reference stars can be satisfied
simultaneously only if the efficiency of \ion{H}{i} collisions is very low
(Sect. 4), few orders of magnitude lower than that given by the Drawin's recipe,
which is consistent with the quantum-mechanical results mentioned above. On
these grounds we do not include inelastic \ion{H}{i} collisions in our reference
model atom.
\subsection{Line selection and atomic data}{\label{sec:atdata}}

In spectra of FGK stars, only a few Sr lines are useful for abundance
determinations. These are two \ion{Sr}{i} and $6$ \ion{Sr}{ii} lines. The
\ion{Sr}{i} resonance line at $4607$ \AA\ is weak, and, thus, sufficiently
reliable for more metal-rich stars ([Fe/H]$ > -1.5$) observed at high
resolution. The two resonance \ion{Sr}{ii} lines, in contrast, remain strong
even in spectra of very metal-poor stars, however, severe blends in the inner
and outer wings lead to systematic errors in abundances, when the blends are not
properly accounted for. The three near-IR \ion{Sr}{ii} lines at $1 \mu$m appear
to be un-blended and are sufficiently strong to be detected even at [Fe/H] $<
-2$.
%
%
%
\begin{table}
\renewcommand{\footnoterule}{}
\renewcommand{\tabcolsep}{2.0pt}
\caption{Lines of \ion{Sr}{i} and \ion{Sr}{ii} selected for abundance calculations.}
\label{lines}
\begin{center}
\begin{tabular}{lrcccr cc}
\hline 
 $\lambda$ & $\Elow$ & low & $\Eup$ & up & $\log gf$ & $\log
(\gamma/N_{\rm H})$\tablefootmark{b} & $\log C_6$\tablefootmark{c} \\
 $\AA$ & [eV] & & [eV] & & & rad cm$^3$ s$^{-1}$ & \\
\hline
\ion{Sr}{i}  & & & & & & & \\ 
 4607.33 & 0.00 & \Sr{5s^2}{1}{S}{}{0}     & 2.69 & \Sr{5p}{1}{P}{\circ}{1} & $ 0.283$  & $-7.53$ & $-31.2$ \\
 7070.01 & 1.85 & \Sr{5p}{3}{P}{\circ}{2}  & 3.60 & \Sr{6s}{3}{S}{}{1}        & $-0.020$ & $-7.15$ & $-30.2$ \\
\ion{Sr}{ii} & & & & & & & \\ 
 4077.71\tablefootmark{a} & 0.00 & \Sr{5s}{2}{S}{}{1/2} & 3.04 & \Sr{5p}{2}{P}{\circ}{3/2} &  $0.158$ & $-7.81$ & $-32.0$ \\
 4167.80                  & 2.94 & \Sr{5p}{2}{P}{\circ}{1/2} & 5.92 & \Sr{6s}{2}{S}{}{1/2} & $-0.502$ & $-7.81$ & $-32.0$ \\
 4215.52\tablefootmark{a} & 0.00 & \Sr{5s}{2}{S}{}{1/2} & 2.94 & \Sr{5p}{2}{P}{\circ}{1/2} & $-0.155$ & $-7.81$ & $-32.0$ \\
 10036.65 & 1.81 & \Sr{4d}{2}{D}{}{3/2} & 3.04 & \Sr{5p}{2}{P}{\circ}{3/2} & $-1.194$ & $-7.64$ & $-31.5$ \\
 10327.31 & 1.84 & \Sr{4d}{2}{D}{}{5/2} & 3.04 & \Sr{5p}{2}{P}{\circ}{3/2} & $-0.240$ & $-7.64$ & $-31.5$ \\
 10914.89 & 1.80 & \Sr{4d}{2}{D}{}{3/2} & 2.94 & \Sr{5p}{2}{P}{\circ}{1/2} & $-0.474$ & $-7.64$ & $-31.5$ \\
\noalign{\smallskip}\hline\noalign{\smallskip}
\end{tabular}
\tablefoot{\\
\tablefoottext{a}{blended}.
\tablefoottext{b}{Log of the FWHM per H atom at $10000$ K}.
\tablefoottext{c}{$C_6$ are in the units of cm$^6$ s$^{-1}$.}
}
\end{center}
\end{table}

%
%
\begin{table}
\renewcommand{\footnoterule}{}
\caption{Hyperfine structure and isotopic shift for the 4077 $\AA$ line.}
\label{hfs}
\begin{center}
\begin{tabular}{lcc}
\hline 
 Isotope & $\lambda$ & $\log gf$ \\
  & $\AA$ & \\
\hline
87  & 4077.697 & $-1.645$ \\
87  & 4077.699 & $-1.485$ \\
84  & 4077.708 & $-2.094$ \\
86  & 4077.709 & $-0.848$ \\
88  & 4077.710 & $0.075$ \\
87  & 4077.724 & $-1.465$ \\
87  & 4077.725 & $-1.956$ \\
\noalign{\smallskip}\hline\noalign{\smallskip}
\end{tabular}
\end{center}
\end{table}

The lines selected in this work are given in Table \ref{lines}. The transition
probabilities were taken from different experimental and theoretical sources.
According to NIST, the \ion{Sr}{i} $gf$-values are very accurate, the
uncertainty is lower than $1$ percent for the $4607$ \AA\ line and better than
$10$ percent for the $7070$ \AA\ subordinate line. These $gf$-values are taken
from the laboratory analysis of \citet{1976RSPTA.281..375P} and
\citet{1988JQSRT..39..477G}, respectively.
The oscillator strengths of the near-IR \ion{Sr}{ii} lines were adopted from the
only available laboratory results of Gallagher (1967). These are fully
consistent with the most recent \textit{ab initio} calculations of
\citet{2002MNRAS.331..875B}, which are also adopted in the NLTE model atom.

Damping widths for the calculation of broadening due to elastic 
collisions with \ion{H}{i} are available for the \ion{Sr}{ii} lines from the
quantum-mechanical calculations of \citet{2000A&AS..142..467B} and for the
\ion{Sr}{i} lines they were kindly provided by the referee. We adopt the values
from Barklem et al. for the near-IR \ion{Sr}{ii} transitions. However, for the
strong resonance \ion{Sr}{ii} lines, which are very sensitive to this parameter,
our spectrum synthesis calculations indicate somewhat lower values, by $\sim 20$
percent. The question is whether this difference can be attributed to the
uncertainties of the theoretical data arising because of certain approximations
in the Anstee, Barklem, and O'Mara (ABO) theory. These include representation of
the interaction potential and collisional dynamics and have been investigated by
\citet{2004JPhB...37..677K} for neutral atoms of Mg, Sr, Ca, and  Na. They find that whereas the semi-classical description of a collision is a sufficiently-good
approximation, the ABO potentials become rather inaccurate at small interatomic separations. However, the effect of the latter is rather to under-estimate the line width that clearly does not explain our finding. For the \ion{Sr}{ii} $4167$ line we assume the same $\gamma$ as for the resonance lines, but note that the line is too weak even in the solar spectrum to be sensitive to this parameter at all.

The adopted values of damping widths at $10000$ K per H atom,
$\gamma/N_{\rm H}$, as well as the commonly-used interaction constants in the
van der Waals-type interaction C$_6$ computed for $T = 5780$ K, are given in the
Table \ref{lines}. Here, the latter is a parameter in the \citet[][Eq.
82,48]{1955QB461.U55} formula needed to reproduce the correct line width $\gamma$ at the given temperature\footnote{Note that C$_6$ is often computed using the Uns\"old (1955, Eq. 82,55) approximation, in which the interaction constant derives from the  effective principal quantum number of lower and upper levels of a transition. However, this approximation is known to yield too small damping constants \citep[e.g.][and references therein]{{2007AIPC..938..111B}}}.

For the $4077$ \ion{Sr}{ii} line, which will be used for the abundance
determinations in Sect. 4, we included the hyperfine structure and
isotopic shifts. The magnetic dipole and electric quadrupole constants were
taken from \citet{1990PhRvC..42.2754B}: A (\Sr{5s^2}{1}{S}{}{0})$= -1000.5$ MHz,
A (\Sr{5p}{2}{P}{\circ}{3/2}) $= -36$ MHz, B (\Sr{5p}{2}{P}{\circ}{3/2}) $=
88.5$ MHz. The data are in agreement with the older experiment by
\citet{1983HyInt..15..177B} and theoretical calculations by Yu et al. (2004).
Isotope shifts for $^{84}$Sr, $^{86}$Sr, $^{87}$Sr were taken from
\citet{1983HyInt..15..177B} and the solar isotopic ratios were
adopted\footnote{http://www.nist.gov/pml/data/comp.cfm}. The wavelengths and
$\log gf$ values of the hyperfine structure (HFS) components are given in Table \ref{hfs}. We note that the 4215 \AA\ \ion{Sr}{i} line is only used for a comparative LTE to NLTE analysis for the reasons discussed in Sect. 3.3.
\section{Statistical equilibrium of Sr}{\label{sec:NLTE}}
The departure coefficients\footnote{The departure coefficients are defined
as the ratio of NLTE to LTE level number densities, $ b_i =
n^{\rm{NLTE}}_i/n^{\rm{LTE}}_i $} for selected levels of Sr are shown as
a function of continuum optical depth $\opd$ at $500$  in Fig. \ref{dep} for
the model atmospheres with parameters corresponding to dwarfs and giants with
[Fe/H] $=0$ and $-2.4$. Only the levels, which give rise to the spectral lines
selected for the abundance analysis (Table \ref{lines}), are included in the
plots. These are two levels of \ion{Sr}{i}, forming the resonance lines at
$4607$ \AA,  and two levels of \ion{Sr}{ii}, which are connected by the $4077$
\AA\ transition commonly used in the analysis of metal-poor stars.
%
\begin{table}
\caption{NLTE abundance corrections for the 4607 Sr I line.}
\label{grid1}
\renewcommand{\tabcolsep}{1.8mm}
\begin{tabular}{lr rrrr rrr}
\hline
$T_\mathrm{eff}$ & [Fe/H]  & \multicolumn{7}{c}{$\Delta_\mathrm{NLTE}$}         \\
$\log g$ &      &  2.20 &   2.60 &   3.00 &   3.40 &   3.80 &   4.20 &   4.60 \\
\hline
 4400 & $-3.0$  &  0.54 &   0.51 &   0.47 & --  &   -- &   --  &   -- \\
 4400 & $-2.4$  &  0.49 &   0.49 &   0.49 & --  &   -- &   --  &   -- \\
 4400 & $-1.2$  &  0.29 &   0.29 &   0.28 & --  &   -- &   --  &   -- \\
 4400 & $-0.6$  &  0.27 &   0.26 &   0.24 & --  &   -- &   --  &   -- \\
 4400 &  $0.0$  &  0.19 &   0.18 &   0.16 & --  &   -- &   --  &   -- \\
 & & & & & & & & \\
 4800 & $-3.0$  &  0.55 &   0.55 &   0.51 &   0.49 &   0.45 &   --   &   --   \\
 4800 & $-2.4$  &  0.50 &   0.51 &   0.51 &   0.49 &   0.46 &   0.43 &   0.36 \\
 4800 & $-1.2$  &  0.32 &   0.33 &   0.33 &   0.33 &   0.32 &   0.30 &   0.28 \\
 4800 & $-0.6$  &  0.17 &   0.17 &   0.18 &   0.18 &   0.17 &   0.16 &   0.14 \\
 4800 &  $0.0$  &  0.19 &   0.18 &   0.18 &   0.17 &   0.15 &   0.13 &   0.10 \\
 & & & & & & & & \\
 5200 & $-3.0$  &  0.50 &   0.52 &   0.52 &   0.50 &   0.47 &   0.45 &   0.42 \\
 5200 & $-2.4$  &  0.47 &   0.48 &   0.49 &   0.48 &   0.45 &   0.43 &   0.41 \\
 5200 & $-1.2$  &  0.34 &   0.35 &   0.36 &   0.36 &   0.35 &   0.34 &   0.32 \\
 5200 & $-0.6$  &  0.19 &   0.20 &   0.20 &   0.20 &   0.20 &   0.19 &   0.17 \\
 5200 &  $0.0$  &  0.09 &   0.10 &   0.10 &   0.11 &   0.11 &   0.10 &   0.09 \\
 & & & & & & & & \\
 5600 & $-3.0$  &  0.44 &   0.45 &   0.46 &   0.46 &   0.45 &   0.42 &   0.40 \\
 5600 & $-2.4$  &  0.40 &   0.42 &   0.43 &   0.43 &   0.43 &   0.40 &   0.38 \\
 5600 & $-1.2$  &  0.31 &   0.32 &   0.33 &   0.33 &   0.33 &   0.32 &   0.31 \\
 5600 & $-0.6$  &  0.23 &   0.24 &   0.24 &   0.24 &   0.24 &   0.23 &   0.21 \\
 5600 &  $0.0$  &  0.09 &   0.10 &   0.10 &   0.10 &   0.11 &   0.10 &   0.09 \\
 & & & & & & & & \\
 6000 & $-3.0$  &  0.39 &   0.39 &   0.39 &   0.39 &   0.39 &   0.37 &   0.35 \\
 6000 & $-2.4$  &  0.35 &   0.36 &   0.36 &   0.37 &   0.36 &   0.35 &   0.34 \\
 6000 & $-1.2$  &  0.27 &   0.28 &   0.28 &   0.28 &   0.28 &   0.28 &   0.27 \\
 6000 & $-0.6$  &  0.22 &   0.23 &   0.23 &   0.23 &   0.23 &   0.22 &   0.21 \\
 6000 &  $0.0$  &  0.14 &   0.14 &   0.14 &   0.14 &   0.13 &   0.13 &   0.12 \\
\hline
\end{tabular}
\end{table}

\begin{table}
\caption{NLTE abundance corrections for the 4077 Sr II line.}
\label{grid2}
\renewcommand{\tabcolsep}{1.3mm}
\begin{tabular}{lr rrrr rrr}
\hline
$T_\mathrm{eff}$ & [Fe/H] & \multicolumn{7}{c}{$\Delta_\mathrm{NLTE}$} \\
$\log g$ &      &  2.20 &   2.60 &   3.00 &   3.40 &   3.80 &   4.20 &   4.60 \\
\hline
 4400 & $-3.9$ &  $-0.07$ &  $-0.04$ &  $-0.04$ &   -- &   --  &  -- &   -- \\
 4400 & $-3.0$ &   0.00 &   0.00 &   0.00 &   -- &   --  &  -- &   -- \\
 4400 & $-2.4$ &   0.00 &   0.00 &   0.00 &   -- &   --  &  -- &   -- \\
 4400 & $-1.2$ &   0.00 &   0.00 &   0.00 &   -- &   --  &  -- &   -- \\
 4400 & $-0.6$ &   0.00 &   0.00 &   0.00 &   -- &   --  &  -- &   -- \\
 4400 & $ 0.0$ &   0.00 &   0.00 &   0.00 &   -- &   --  &  -- &   -- \\
 & & & & & & & & \\
 4800 & $-3.9$ &  $-0.11$ &  $-0.08$ &  $-0.04$ &  $-0.01$ &  $-0.01$  &  --   &   --   \\
 4800 & $-3.0$ &  $-0.02$ &  $-0.02$ &  $-0.02$ &  $-0.01$ &  $-0.01$  & $-0.01$ &   0.00 \\
 4800 & $-2.4$ &   0.00 &   0.00 &   0.00 &   0.00 &  $-0.01$  & $-0.01$ &   0.00 \\
 4800 & $-1.2$ &   0.00 &   0.00 &   0.00 &   0.00 &   0.00  &  0.00 &   0.00 \\
 4800 & $-0.6$ &   0.00 &   0.00 &   0.00 &   0.00 &   0.00  &  0.00 &   0.00 \\
 4800 & $ 0.0$ &   0.00 &   0.00 &   0.00 &   0.00 &   0.00  &  0.00 &   0.00 \\
 & & & & & & & & \\
 5200 & $-3.9$ & $-0.07$ &  $-0.03$ &   0.01 &   0.05 &   0.08  &  0.12 &   0.13 \\
 5200 & $-3.0$ & $-0.05$ &  $-0.06$ &  $-0.06$ &  $-0.05$ &  $-0.04$  & $-0.03$ &  $-0.01$ \\
 5200 & $-2.4$ & $-0.01$ &  $-0.01$ &  $-0.01$ &  $-0.01$ &  $-0.01$  & $-0.01$ &  $-0.01$ \\
 5200 & $-1.2$ &   0.00 &   0.00 &   0.00 &   0.00 &  $-0.01$  & $-0.01$ &  $-0.01$ \\
 5200 & $-0.6$ &   0.00 &   0.00 &   0.00 &   0.00 &   0.00  &  0.00 &   0.00 \\
 5200 & $ 0.0$ &   0.00 &   0.00 &   0.00 &   0.00 &   0.00  &  0.00 &   0.00 \\
 & & & & & & & & \\
 5600 & $-3.9$ &   0.05 &   0.08 &   0.11 &   0.13 &   0.14  &  0.14 &   0.15 \\
 5600 & $-3.0$ &  $-0.07$ &  $-0.08$ &  $-0.09$ &  $-0.08$ &  $-0.06$  & $-0.04$ &  $-0.01$ \\
 5600 & $-2.4$ &  $-0.01$ &  $-0.03$ &  $-0.03$ &  $-0.04$ &  $-0.04$  & $-0.04$ &  $-0.03$ \\
 5600 & $-1.2$ &   0.00 &   0.00 &  -0.01 &  -0.01 &  -0.01  & -0.01 &  -0.01 \\
 5600 & $-0.6$ &   0.00 &   0.00 &   0.00 &  -0.01 &  -0.01  & -0.01 &  -0.01 \\
 5600 & $ 0.0$ &   0.00 &   0.00 &   0.00 &   0.00 &   0.00  &  0.00 &   0.00 \\
 & & & & & & & & \\
 6000 & $-3.9$ &   0.19 &   0.19 &   0.19 &   0.18 &   0.17  &  0.16 &   0.15 \\
 6000 & $-3.0$ &  $-0.04$ &  $-0.06$ &  $-0.06$ &  $-0.05$ &  $-0.04$  & $-0.02$ &   0.01 \\
 6000 & $-2.4$ &  $-0.01$ &  $-0.03$ &  $-0.05$ &  $-0.06$ &  $-0.07$  & $-0.07$ &  $-0.06$ \\
 6000 & $-1.2$ &   0.01 &   0.00 &  $-0.01$ &  $-0.01$ &  $-0.01$  & $-0.01$ &  $-0.01$ \\
 6000 & $-0.6$ &   0.00 &   0.00 &   0.00 &  $-0.01$ &  $-0.01$  & $-0.01$ &  $-0.01$ \\
 6000 &  $0.0$ &   0.00 &   0.00 &   0.00 &   0.00 &   0.00  &  0.00 &   0.00 \\
\hline
\end{tabular}
\end{table}

\begin{table}
\caption{NLTE abundance corrections for the 10036 Sr II line.}
\label{grid3}
\renewcommand{\tabcolsep}{1.3mm}
\begin{tabular}{lr rrrr rrr}
\hline
$T_\mathrm{eff}$ & [Fe/H]  & \multicolumn{7}{c}{$\Delta_\mathrm{NLTE}$} \\
$\log g$ &      &  2.20 &   2.60 &   3.00 &   3.40 &   3.80 &   4.20 &   4.60 \\
\hline
 4400 & $-3.9$ & $-0.14$ & $ -0.15$ &  $-0.18$ &  -- &  -- &  -- &  -- \\
 4400 & $-3.0$ & $-0.13$ & $ -0.12$ &  $-0.12$ &  -- &  -- &  -- &  -- \\
 4400 & $-2.4$ & $-0.13$ & $ -0.12$ &  $-0.12$ &  -- &  -- &  -- &  -- \\
 4400 & $-1.2$ & $-0.17$ & $ -0.15$ &  $-0.13$ &  -- &  -- &  -- &  -- \\
 4400 & $-0.6$ & $-0.15$ & $ -0.14$ &  $-0.13$ &  -- &  -- &  -- &  -- \\
 4400 & $ 0.0$ & $-0.10$ & $ -0.11$ &  $-0.10$ &  -- &  -- &  -- &  -- \\
       & & & & & & & & \\
 4800 & $-3.9$ & $-0.18$ & $ -0.18$ &  $-0.19$ &  $-0.20$ &  $-0.12$ &  --  &  -- \\
 4800 & $-3.0$ & $-0.13$ & $ -0.13$ &  $-0.12$ &  $-0.12$ &  $-0.12$ &  $-0.09$ &  $-0.08$ \\
 4800 & $-2.4$ & $-0.15$ & $ -0.14$ &  $-0.13$ &  $-0.12$ &  $-0.10$ &  $-0.09$ &  $-0.05$ \\
 4800 & $-1.2$ & $-0.19$ & $ -0.17$ &  $-0.15$ &  $-0.12$ &  $-0.09$ &  $-0.07$ &  $-0.06$ \\
 4800 & $-0.6$ & $-0.17$ & $ -0.16$ &  $-0.14$ &  $-0.12$ &  $-0.09$ &  $-0.07$ &  $-0.05$ \\
 4800 & $ 0.0$ & $-0.11$ & $ -0.12$ &  $-0.11$ &  $-0.10$ &  $-0.08$ &  $-0.07$ &  $-0.05$ \\
       & & & & & & & & \\
 5200 & $-3.9$ & $-0.19$ & $ -0.19$ &  $-0.19$ &  $-0.19$ &  $-0.17$ &  $-0.15$ &  $-0.13$ \\
 5200 & $-3.0$ & $-0.13$ & $ -0.12$ &  $-0.12$ &  $-0.12$ &  $-0.11$ &  $-0.10$ &  $-0.09$ \\
 5200 & $-2.4$ & $-0.14$ & $ -0.14$ &  $-0.13$ &  $-0.12$ &  $-0.10$ &  $-0.09$ &  $-0.08$ \\
 5200 & $-1.2$ & $-0.20$ & $ -0.19$ &  $-0.16$ &  $-0.13$ &  $-0.10$ &  $-0.08$ &  $-0.06$ \\
 5200 & $-0.6$ & $-0.19$ & $ -0.18$ &  $-0.16$ &  $-0.13$ &  $-0.10$ &  $-0.07$ &  $-0.05$ \\
 5200 & $ 0.0$ & $-0.13$ & $ -0.13$ &  $-0.12$ &  $-0.11$ &  $-0.09$ &  $-0.07$ &  $-0.05$ \\
       & & & & & & & & \\
 5600 & $-3.9$ & $-0.16$ & $ -0.15$ &  $-0.14$ &  $-0.12$ &  $-0.10$ &  $-0.06$ &  $-0.03$ \\
 5600 & $-3.0$ & $-0.10$ & $ -0.10$ &  $-0.10$ &  $-0.10$ &  $-0.09$ &  $-0.07$ &  $-0.05$ \\
 5600 & $-2.4$ & $-0.12$ & $ -0.11$ &  $-0.10$ &  $-0.09$ &  $-0.08$ &  $-0.07$ &  $-0.06$ \\
 5600 & $-1.2$ & $-0.20$ & $ -0.19$ &  $-0.17$ &  $-0.14$ &  $-0.11$ &  $-0.08$ &  $-0.06$ \\
 5600 & $-0.6$ & $-0.19$ & $ -0.19$ &  $-0.17$ &  $-0.15$ &  $-0.12$ &  $-0.08$ &  $-0.06$ \\
 5600 & $ 0.0$ & $-0.14$ & $ -0.14$ &  $-0.14$ &  $-0.13$ &  $-0.11$ &  $-0.08$ &  $-0.06$ \\
      & & & & & & & & \\
  6000 & $-3.9$ & $-0.08$ & $ -0.06$ &  $-0.04$ &  $-0.02$ &  $ 0.01$ &  $ 0.04$ &  $ 0.07$ \\
 6000 & $-3.0$ & $-0.06$ & $ -0.06$ &  $-0.05$ &  $-0.05$ &  $-0.03$ &  $-0.02$ &  $ 0.01$ \\
 6000 & $-2.4$ & $-0.08$ & $ -0.07$ &  $-0.06$ &  $-0.06$ &  $-0.05$ &  $-0.04$ &  $-0.03$ \\
 6000 & $-1.2$ & $-0.17$ & $ -0.16$ &  $-0.14$ &  $-0.12$ &  $-0.10$ &  $-0.08$ &  $-0.06$ \\
 6000 & $-0.6$ & $-0.18$ & $ -0.18$ &  $-0.17$ &  $-0.15$ &  $-0.12$ &  $-0.09$ &  $-0.06$ \\
 6000 & $ 0.0$ & $-0.14$ & $ -0.15$ &  $-0.14$ &  $-0.14$ &  $-0.12$ &  $-0.09$ &  $-0.07$ \\
\hline
\end{tabular}
\end{table}

\subsection{NLTE effects in \ion{Sr}{i}}\label{sec:sri}
\begin{figure*}
\begin{center}
\hbox{\resizebox{\columnwidth}{!}
{{\includegraphics{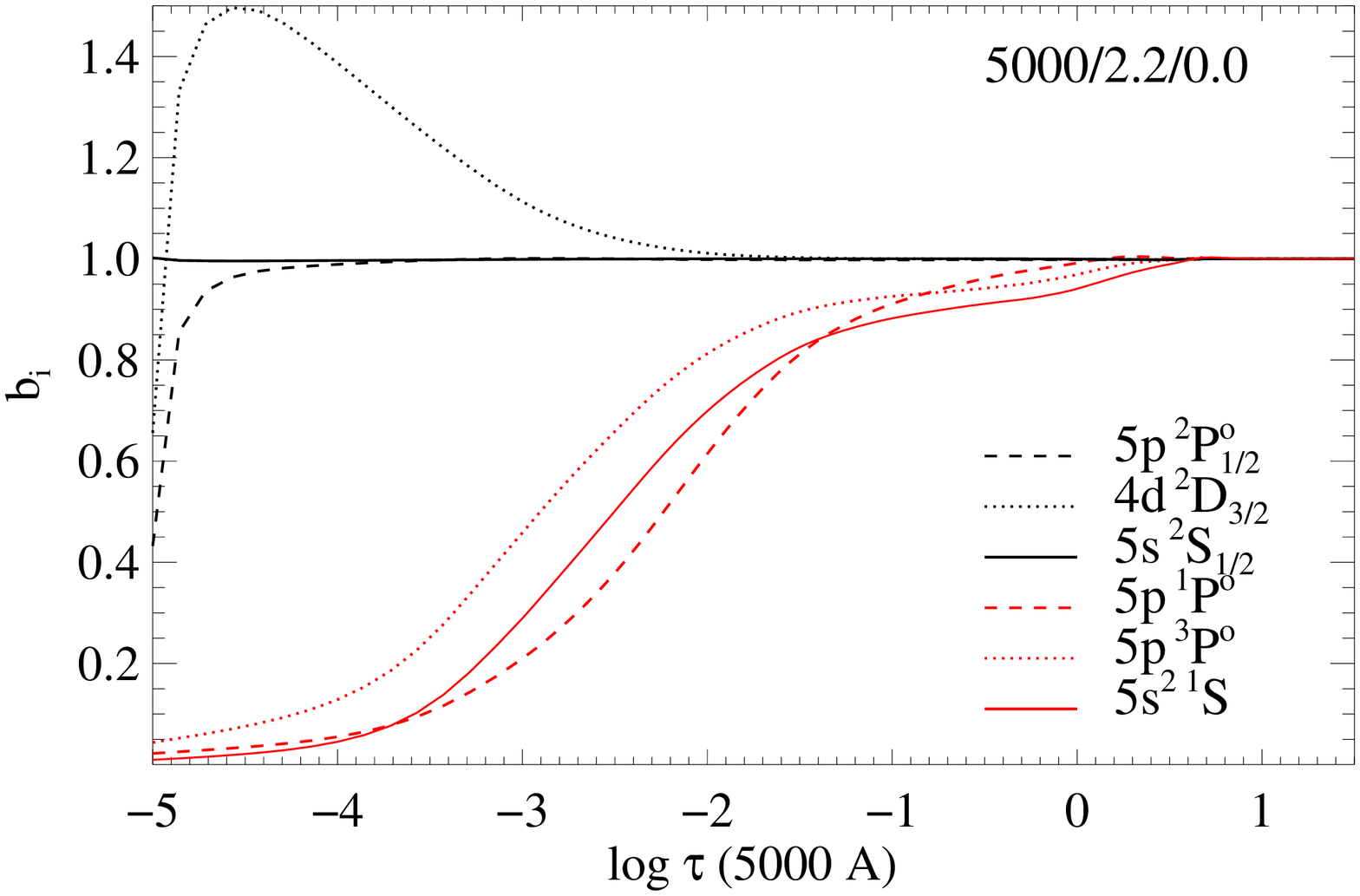}}} \hfill
\resizebox{\columnwidth}{!}
{{\includegraphics{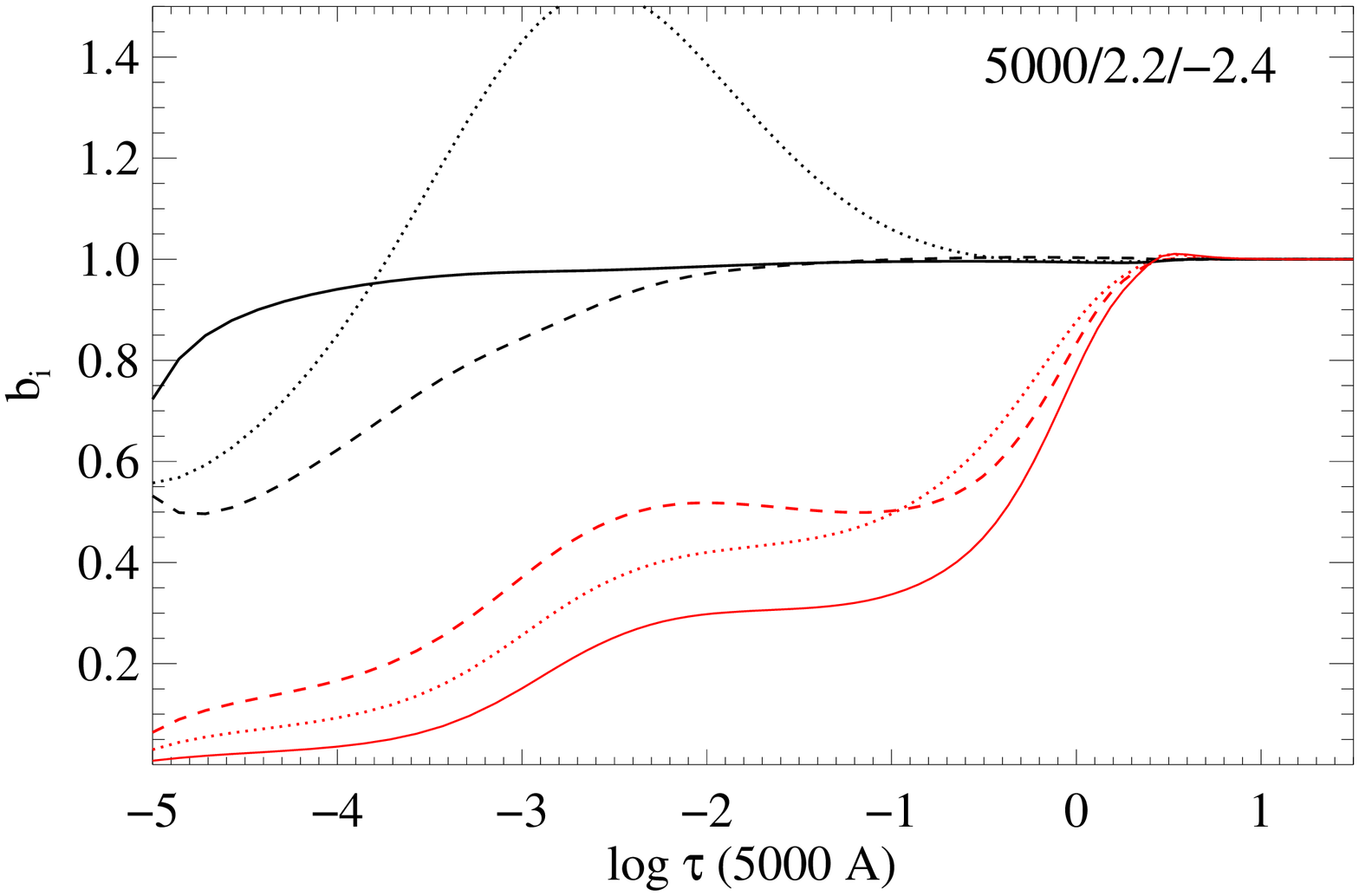}}}}
\hbox{\resizebox{\columnwidth}{!}
{{\includegraphics{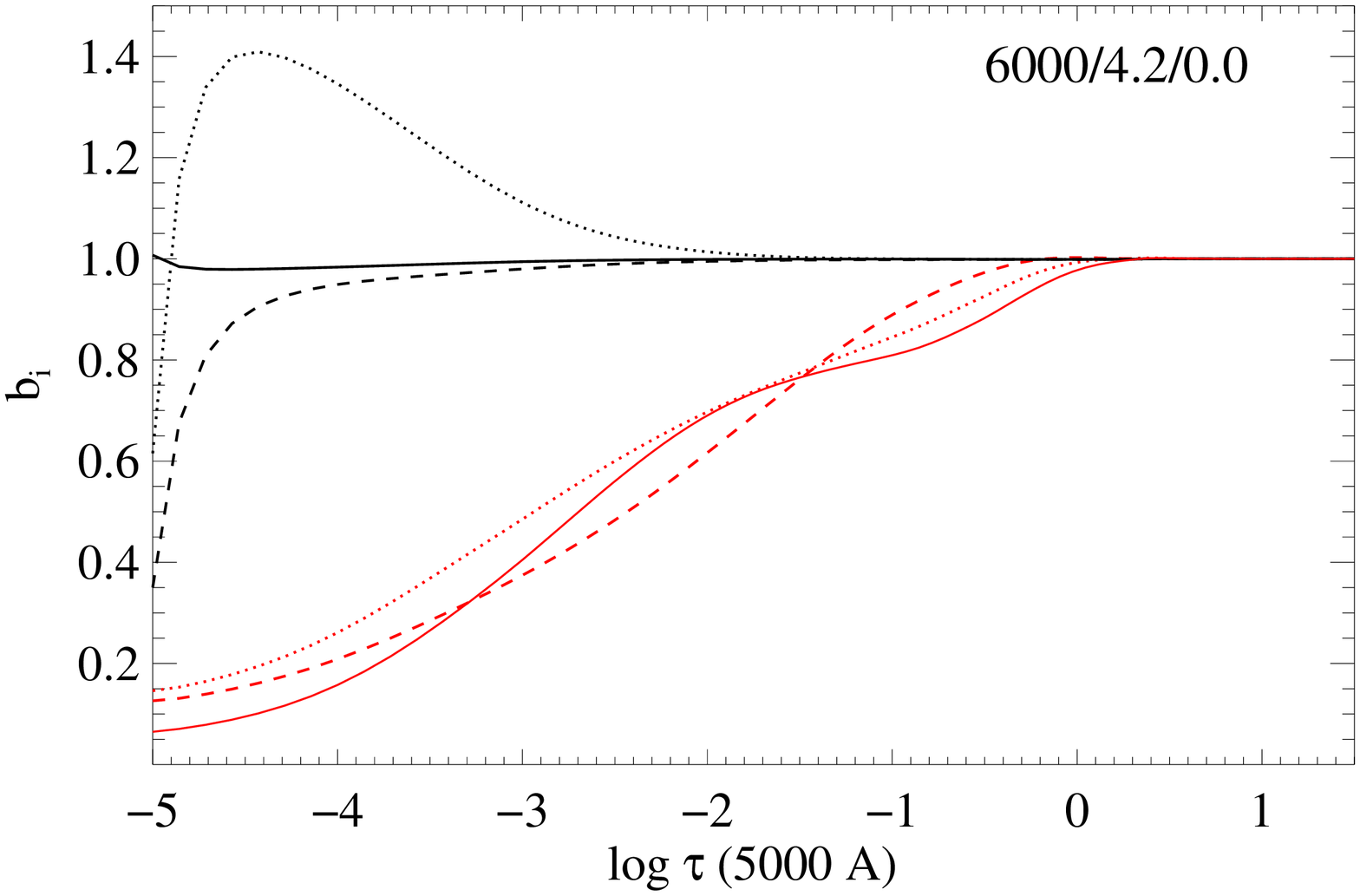}}} \hfill
\resizebox{\columnwidth}{!}
{{\includegraphics{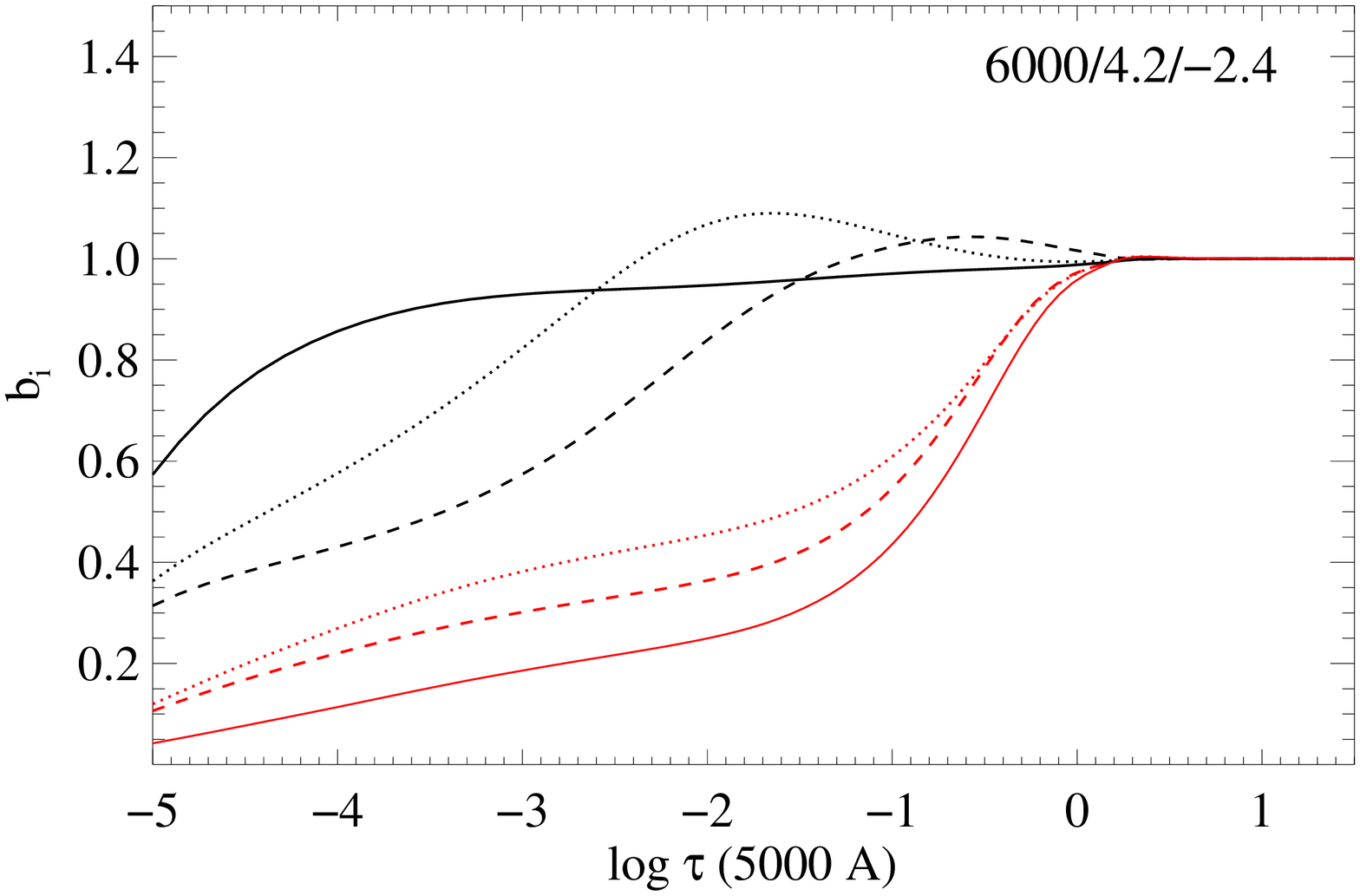}}}}
\caption{Departure coefficients $b_i$ of selected \ion{Sr}{i} and \ion{Sr}{ii}
levels for different stellar parameters, $\Teff$, $\log g$, and [Fe/H] (indicated in each sub-plot). Red: \ion{Sr}{i} levels, \Sr{5s^2}{1}{S}{}{} (solid),
\Sr{5p}{3}{P}{\circ}{} (dotted), \Sr{5p}{1}{P}{\circ}{} (dashed). Black:
\ion{Sr}{ii} levels, \Sr{5s}{2}{S}{}{1/2} (solid), \Sr{4d}{2}{D}{}{3/2}
(dotted), \Sr{5p}{2}{P}{\circ}{1/2} (dashed).}
\label{dep}
\end{center}
\end{figure*}
\ion{Sr}{i} with the ionization potential $5.69$ eV is a trace atom in the
the atmospheres of late-type stars. The ratio of \ion{Sr}{i}/\ion{Sr}{ii} falls
below $10^{-3}$ above continuum optical depth unity, and the departures from
LTE in the distribution of atomic level populations are almost entirely due to
overionization. The behavior of the \ion{Sr}{i} departure coefficients with
stellar parameters, thus, resembles that of the similar trace atoms, such as
\ion{Co}{i} \citep{2010MNRAS.401.1334B} or \ion{Cr}{i}
\citep{2010A&A...522A...9B}.
The levels are underpopulated and their $b_i$-factors monotonously decrease
outwards. The ground state \Sr{5s^2}{1}{S}{}{} with the ionization edge at
$2177$ \AA\ is also strongly over-ionized and it decouples from the other
low-excited levels because collisions can not bridge the large energy gap of
$\sim 2$ eV between them. Line transitions have a small influence on the
statistical equilibrium of \ion{Sr}{i} causing some changes of the level
populations only in the solar-metallicity models. For example, in the metal-rich
models (Fig. \ref{dep}, left panel), the departure coefficient of the 5p
\element[][1]{P^o} level drops below that of the 5s$^2$ \element[][1]{S} at
$\opd \approx -1$, marking the depth at which the $4607$ resonance line becomes
optically thin, and spontaneous transitions depopulate the upper level of the
transition. 

In the metal-poor models (Fig. \ref{dep}, right panel), the coupling of the
levels by radiative bound-bound transitions and collisions
diminishes\footnote{Note that we do not include the \ion{H}{i} collisions as
discussed in Sect. 4.} even though the collision rates between the low-excited
levels are large enough to keep them are in detailed
balance even in the models with [Fe/H] $= -1$. The populations are now set
primarily by the balance between radiative ionizations and recombinations,
and the run of $b_i$-factors with $\opd$ simply reflects the mean intensity vs
Planck function inequalities, $J_\nu \neq B_\nu$, at the frequencies of the
level photoionization continua at each depth point. For example, in the model
with $\Teff = 6000$ K, $\log g = 4.2$, and [Fe/H] $= -2.4$, the lowest energy
levels 5p \element[][3]{P^o} and 5p \element[][1]{P^o} decouple from each other
and from the continuum at $\opd \approx 0$. Then, in a very narrow interval of
optical depths, $-1 < \opd < 0$, their departure coefficients demonstrate a
sudden drop following the steep temperature gradient of the atmosphere, and
thus, rapidly increasing $J_\nu - B_\nu$ imbalance (note that matter becomes
transparent to the radiation above ionization edges of both levels, $3202$ and
$4126$ \AA\ at $\opd = +0.11$ respectively $\opd = +0.05$). Further out, $\opd <
-1$, where the mean intensity as well as the local kinetic temperature remain
roughly constant, the departure coefficients smooth out, slowly decreasing
outwards.

The NLTE effects on the formation of the \ion{Sr}{i} line at $4607$ \AA\
are generally to decrease the line opacity, shifting the line $\tau_\nu$ scale
to the deeper hotter atmospheric layers and, thus, leading to weaker lines
compared to the LTE case.
\subsection{NLTE effects in \ion{Sr}{ii}}\label{sec:srii}
The NLTE effects on the \ion{Sr}{ii} levels are due to the non-equilibrium
excitation in the line transitions. This was extensively investigated by
\citet{1997AZh....74..601B} and \citet{2006ApJ...641..494S}, and our results
are qualitatively very similar to these studies. The analysis of radiative and
collisional rates populating and de-populating the levels, as well as trial
calculations with atomic models devoid of some key transitions, in particular
the resonance and subordinate lines, allows us to draw the following
conclusions.

In the solar-metallicity models (Fig. \ref{dep}, left panel), the ground state,
\Sr{5s}{2}{S}{}{} ($\lambda_{\rm thr} \sim 1130 \AA$) and the lowest odd level
\Sr{5p}{2}{P}{\circ}{} level ($\lambda_{\rm thr} \sim 1540 \AA$ , $E \sim 3$ eV)
have nearly LTE populations out to $\opd \sim -3$, and become underpopulated only in the outermost layers, which are transparent to the radiation across their ionization edges. Still, there is small net radiative excitation in the transitions from
these levels upwards, which very efficiently populates the levels with
excitation energy $E \geq 6$ eV, such as \Sr{5d}{2}{D}{}{},
\Sr{6p}{2}{P}{\circ}{}, and \Sr{7s}{2}{S}{}{} (not shown in the plots) due to
their small Boltzmann factors. The high-excited levels are, however, also
over-ionized, which in turn, increases the number density of the \ion{Sr}{iii}
ground state. The lowest metastable \ion{Sr}{ii} level \Sr{4d}{2}{D}{}{} ($E
\sim 1.8$ eV) is nearly in thermal equilibrium with \Sr{5p}{2}{P}{\circ}{} due
to the dominance of collision rates in the deeper layers, however in the outer
layers it gains appreciable overpopulation due to the photon losses in the
near-IR transitions between the two levels. In fact, this process is a part of
photon suction in a sequence of low-frequency transitions, which connect
\Sr{4d}{2}{D}{}{} to a large number of high-levels close to the continuum. Many
of these transitions become optically thin at $\opd < -3$, so that spontaneous
de-excitation lead to the overpopulation \Sr{4d}{2}{D}{}{}, the lowest energy
state of the cascading sequence.

The departures from LTE change with decreasing metallicity (Fig. \ref{dep},
right panel). First, the line transitions weaken and and over-population of
\Sr{4d}{2}{D}{}{} occurs in the deeper layers compared to the solar metallicity
models. Also, radiative pumping in the resonance transitions at $4077$ and
$4215$ \AA\ is more efficient due to larger radiative fluxes. This process leads
to a
marked over-population of the \Sr{5p}{2}{P}{\circ}{} level at $-1 < \opd < 0$,
especially in the hotter model, $\Teff = 6200$ K. Outside these optical depths,
over-ionization dominates and all low-lying levels become underpopulated. The
importance of strong line scattering was tested by excluding the resonance lines
and subordinate lines at $1$ $\mu$m from the model atom. The overpopulation of
the \Sr{5p}{2}{P}{\circ}{} and \Sr{4d}{2}{D}{}{} vanished both for the
metal-rich and metal-poor models.
\begin{figure*}
\begin{center}
\resizebox{0.9\linewidth}{!}{\includegraphics{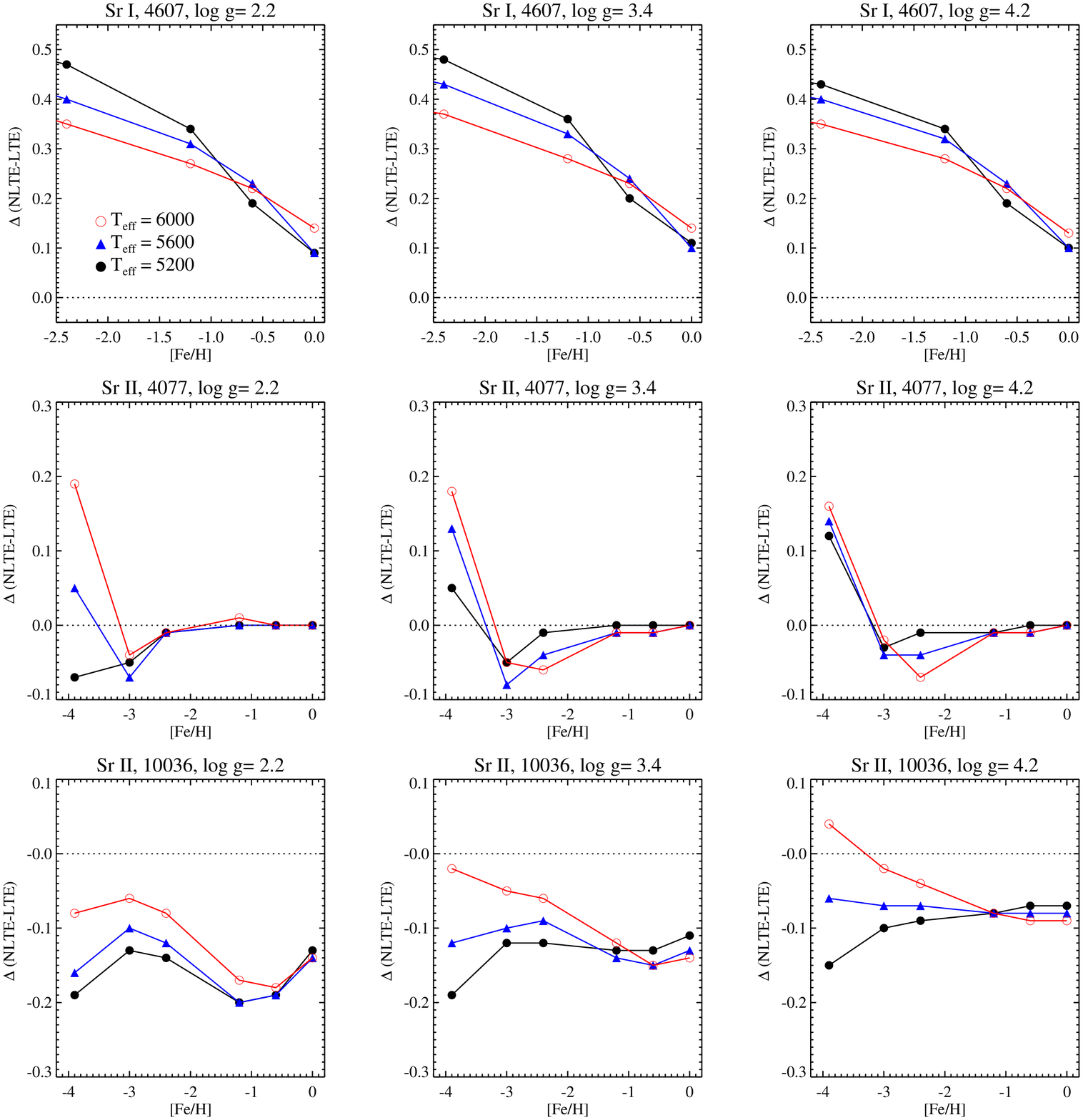}}
\caption{NLTE abundance corrections for the \ion{Sr}{i} and \ion{Sr}{ii} lines.
Note different x-axis scales.}
\label{nlteab}
\end{center}
\end{figure*}
\subsection{NLTE abundance corrections}\label{sec:abuc}
To quantify the effect of NLTE on the abundance determinations, we use the
concept of the NLTE abundance correction $\drs$, where:
\begin{equation}
\drs = \log \epsilon \rm{(Sr)}_{\rm NLTE} - \log \epsilon \rm{(Sr)}_{\rm LTE}
\end {equation}
is the the logarithmic correction, which has to be applied to an LTE abundance
determination of a specific line to obtain the exact value corresponding to the
use of NLTE line formation. These values were computed for a grid of model
atmospheres by equalizing the NLTE equivalent widths though varying the element
abundance in the LTE calculations. The NLTE abundance corrections for the
$4077$, $4607$ and $10036$ line are given in Tables \ref{grid1}, \ref{grid2},
and \ref{grid3} for a wide range of stellar parameters. Fig. \ref{nlteab} (top
panel) also shows the values of $\drs$ for selected model atmospheres as a
function of effective temperature, gravity, and [Fe/H]. 

As evident from Fig. \ref{nlteab} (top panel) and Table \ref{grid1}, the NLTE
abundance corrections for the $4607$ \AA\ are always positive, and they are
maximum for cool giants with sub-solar metallicity, [Fe/H] $< -1$. This behavior
can be easily understood by inspecting the plots of departure coefficients (Fig.
\ref{dep}, top right panel): at the typical depths of the resonance line
formation, $\opd \sim 0 ... -0.5$, the level populations are
already depleted by a factor of two compared to LTE. In the models of warmer
metal-poor dwarfs, the line formation is confined to the deeper layers, $\opd
\sim 0$, where the NLTE effects in the \ion{Sr}{i} ionization balance are
smaller. This is reflected in the behavior of the abundance corrections, which 
also decrease. $\drs$ are comparatively small for the warmest models,
corresponding to horizontal branch stars ($\Teff = 6200$ and $\log g = 2.2$).
Note, however, that at $\Teff > 6000$ K and [Fe/H]$< -1.5$ the $4607$ \AA\ line
is very weak, $\EW < 0.5$ \mA, and, thus, is not useful for abundance
determinations. In contrast, the line could be suitable for the analysis in
sufficiently good-quality (R $>$ 20000, S/N $>$50) spectra of giants and
sub-giants down to [Fe/H] $\sim -2.5$.

As described in the previous section, the formation of \ion{Sr}{ii} lines is
also significantly affected by the NLTE effects. Inspection of Table \ref{grid2}
and Fig. \ref{nlteab} (middle panel) shows that for the resonance \ion{Sr}{ii}
line at $4077$ \AA\ the NLTE abundance corrections are typically negative for
stars with metallicity [Fe/H]$> -3$, but rapidly increase for more metal-poor
stars with $\Teff > 5000$ K and/or $\log g > 3$. The amplitude of $\drs$
increases with $\Teff$ and decreasing $\log g$. Another resonance \ion{Sr}{ii}
line at $4215$ \AA\ shows the same pattern and the NLTE abundance corrections
are nearly identical; we caution, however, that this line is blended by an
\ion{Fe}{i} line, and should be avoided in abundance analyses of cool stars. The
implication is that, compared to NLTE, traditional LTE studies relying on the resonance \ion{Sr}{ii} lines under-estimate the abundance of Sr in very metal-poor stars, but somewhat over-estimate it for less metal-poor objects, such at that of thick and thin disk of the Galaxy.

For the near-IR subordinate \ion{Sr}{ii} lines, the NLTE abundance corrections
are typically negative (Table \ref{grid3} and Fig. \ref{nlteab}, bottom panel),
being as large as $-0.2$ dex even for mildly metal-poor giants; this behaviour
is consistent with the discussion in Sect. \ref{sec:srii}, as the NLTE effects
are due to strong line scattering. However, $\drs$ are usually within $-0.1$ dex
for dwarfs and become mildly positive only at very low metallicity, [Fe/H] $<
-3$ and $\Teff > 6000$ K. The NLTE abundance corrections presented in Table
\ref{grid3} are rather similar for the other two members of the near-IR triplet,
i.e., the lines at $10327$ and $10914$ \AA, and can be provided by request.

Our NLTE abundance corrections are in line with the results of
\citet{2006ApJ...641..494S}, although they investigated the formation of the
resonance \ion{Sr}{ii} lines only. They find NLTE line strengthening for the
giant models with [Fe/H] $= -1 \ldots -2$ and NLTE weakening for the models with
[Fe/H] $= -4 \ldots -5$. However, we note some qualitative differences with \citet{2011A&A...530A.105A}. For example, for the near-IR \ion{Sr}{ii} lines their abundance corrections at [Fe/H] $= -3$ and $\Teff = 4800 - 5300$ K are of the order of $-0.3$ to $-0.5$ dex and are very sensitive to surface gravity (their Fig. 7, top panel). In
contrast, our values for the $10036$ \AA\ line remain within $\sim -0.1$ dex for any
$\log g$ value in this [Fe/H] and $\Teff$ range. For the resonance \ion{Sr}{ii}
line at  $4077$ \AA, our $\drs$ are mildy negative for any $\log g$ and $\Teff$
at [Fe/H] $=-3$, whereas \citet{2011A&A...530A.105A} obtain large positive
corrections for dwarfs and negative $\drs$ for giants. The differences are most
likely caused by the differences in the NLTE model atoms, as described in Sect. \ref{sec:modelatom}.

\section{Application to the abundance analysis of late-type stars}
\begin{figure}
\centering
\includegraphics[scale=0.8]{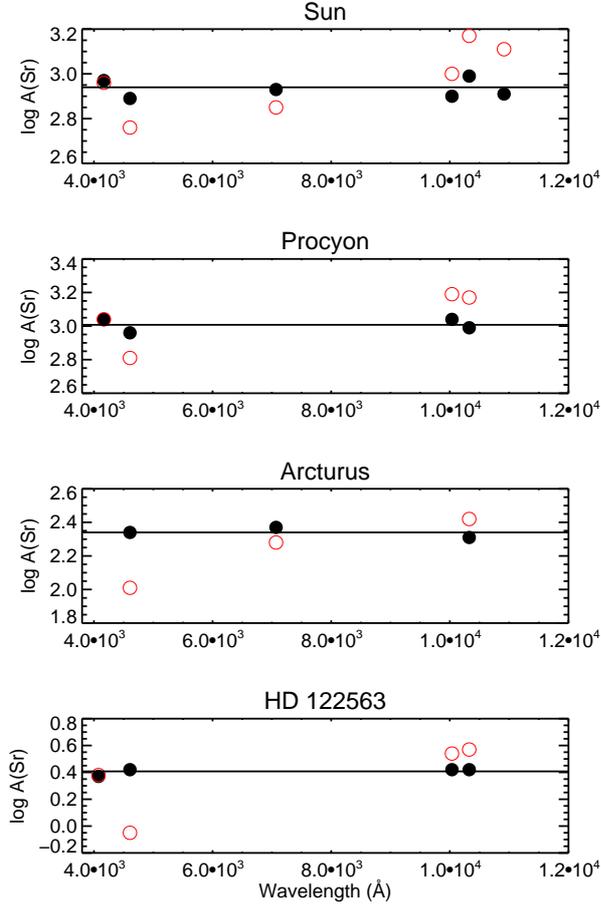}
\caption{NLTE and LTE line-by-line Sr abundances for the selected stars. In
contrast to LTE (red open circles), the \ion{Sr}{i} and \ion{Sr}{ii} lines
provide consistent abundances in NLTE (black filled circles).}
\label{abundances}
\end{figure}

In addition to the Sun, three other late-type stars were selected for the abundance analysis. The principle selection criteria were that stellar parameters are well-constrained by independent techniques (interferometry, parallaxes), observations cover a large wavelength interval and their quality is very high for the 4607 \ion{Sr}{i} line and the near-IR \ion{Sr}{ii} triplet to be accurately measured. Stellar parameters for the warm turn-off star Procyon, moderately metal-poor giant Arcturus (HD 124897), and very metal-poor giant HD 122563 were taken from \citet{2003A&A...407..691K}, \citet{2009A&A...504..543T}, and \citet{2008A&A...478..529M}, respectively (Table \ref{sample}). The accuracy of these parameters was verified in our recent analysis of Fe NLTE statistical equilibrium with mean 3D model atmospheres \citep{bergemann12}. We stress that it is not our goal here to obtain Sr abundances for a large sample of stars, as this will be the focus of our forthcoming publications (Hansen et al. in prep., Ruchti et al. in prep.), but rather to test the performance of the new NLTE model atom.

For the spectrum synthesis (Sect. 2.1), we used the solar KPNO flux spectrum
(Kurucz et al. 1984). The spectra of \object{Arcturus} (HD 124897),
\object{Procyon} (\object{HD 61421}), and \object{HD 122563} were retrieved from
the UVESPOP database \citep{2003Msngr.114...10B}. These UVES (VLT, Paranal)
spectra have a slit-determined resolution of $\sim 80000$ and a signal-to-noise
ratio $S/N \sim 300$ near 5000 \AA.

The solar abundance was determined using five Sr lines from the Table
\ref{lines}, adopting $\Teff = 5777$, $\log g = 4.44$, [Fe/H] $=0$, and $\Vmic
= 1$ km/s. The results are largely discrepant in LTE: $\log \epsilon_{\rm Sr I}
= 2.81 \pm 0.06$ and $\log \epsilon_{\rm Sr II} = 3.06 \pm 0.1$. The NLTE model
atom successfully eliminates this problem. The abundances from the two
ionization stages are fully consistent $\log \epsilon_{\rm Sr I} = 2.91 \pm
0.03$ and $\log \epsilon_{\rm Sr II} = 2.94 \pm 0.04$ in NLTE. Here, the error
corresponds to one standard deviation of the line sample. The NLTE result is
also in agreement with the meteoritic abundance of Sr, $\log \epsilon =
2.90 \pm 0.03$ dex \citep{2009LanB...4B...44L}. We note that the meteoritic
abundance of Sr given by \citet{2009ARA&A..47..481A}, Sr, $\log \epsilon
= 2.88 \pm 0.03$ dex, is slightly lower, which is a consequence of different
reference Si abundances. The solar Sr abundance was also determined by
\citet[][]{2000MNRAS.311..535B}, \citet{2011A&A...530A.105A}, and Mashonkina \&
Gehren (2001). The last two references perform NLTE calculations, which are
generally in agreement with our results, both in terms of NLTE abundance
corrections and absolute abundances. We caution, however, that this agreement is
not particularly telling, because other parameters in the modeling are
different. Mashonkina \& Gehren (2001) adopted $\Vmic = 0.8$ km/s and employed
the older (ODF) version of the MAFAGS model atmospheres.
\citet{2000MNRAS.311..535B} recover meteoritic Sr abundance from the neutral Sr
lines even in LTE, which is not unexpected as they used $\Vmic = 0.85$ km/s and
the solar Holweger-M\"uller (1974) model atmosphere\footnote{The
semi-empirical Holweger-M\"uller model atmosphere is hotter than all solar
theoretical models by $\sim$ 150 K over the line formation layers and, in this
way, it 'mimics' NLTE effects for some atoms}.
%
%
%
\begin{table*}
\begin{center}
\caption[]{Basic parameters and the Sr abundances for the selected stars. }
\label{sample}
\begin{tabular}{l l l l c l l l l }
\hline
\hline
Star  & T$_{\rm eff}$ & $\log g$ & [Fe/H] & $\xi $ &
 \multicolumn{2}{c}{LTE} & \multicolumn{2}{c}{NLTE} \\
      & K             &  (cgs)   &        & km/s   &
  [\ion{Sr}{i}/Fe] & [\ion{Sr}{ii}/Fe] & [\ion{Sr}{i}/Fe] & [\ion{Sr}{ii}/Fe] \\   
\hline
HD 61421    & 6510  & 3.96    & $-0.03$  & 1.8  & $-0.08$   & ~~~$0.27$  & ~~~0.07     & ~~~0.15 \\
HD 122563   & 4600  & 1.60    & $-2.50$  & 1.8  & $-0.47$   & ~~~$0.08$ & ~~~$0.00$ & $-0.02$  \\
HD 124897   & 4300  & 1.50    & $-0.50$  & 1.5  & $-0.28$   & ~~~$0.00$ & $-0.07$       & $-0.11$  \\
\hline
\hline
\end{tabular}
\end{center}
\end{table*}

\begin{figure*}
\centering
\hbox{
\resizebox{\columnwidth}{!}{\rotatebox{0}
{\includegraphics{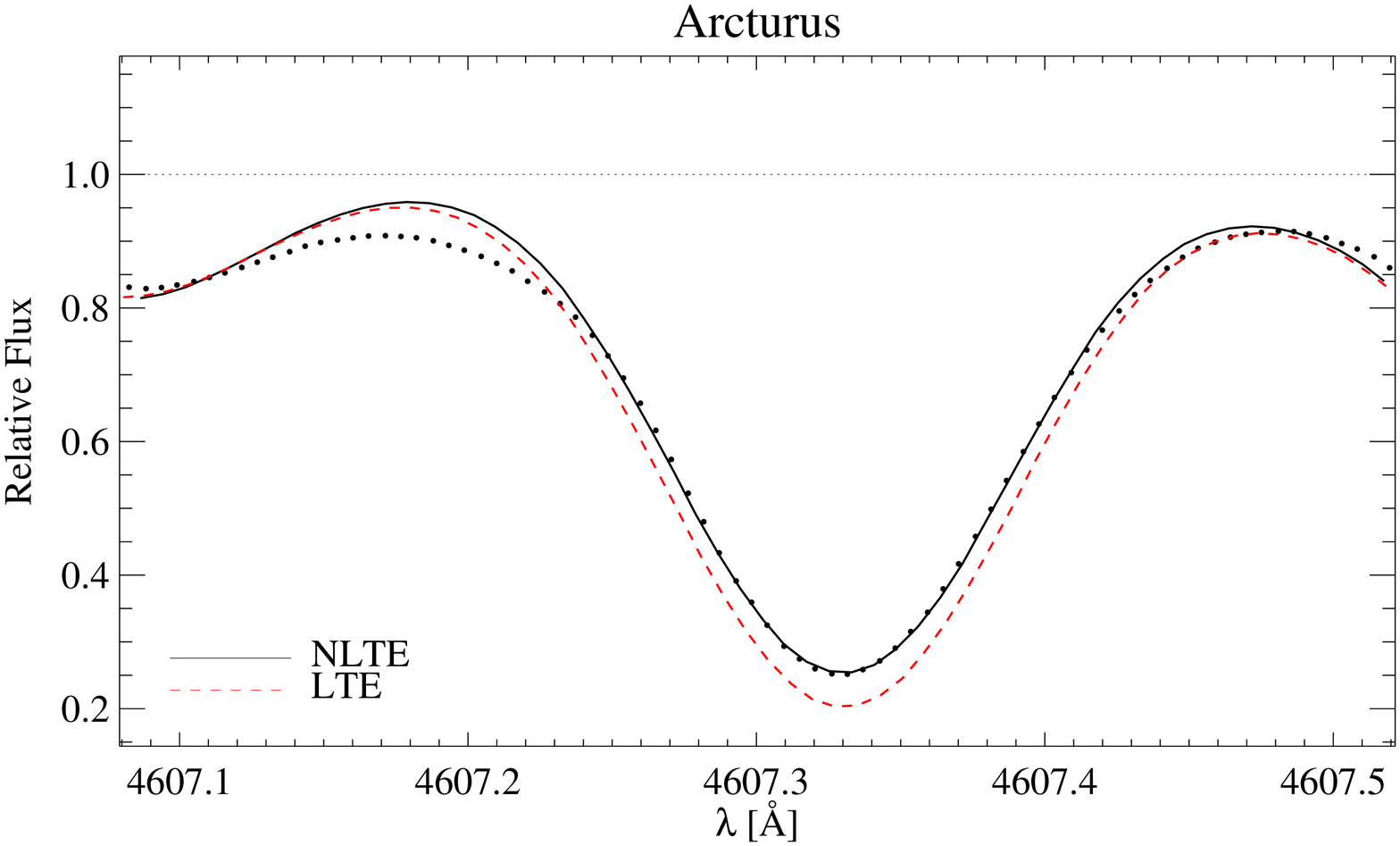}}}\hfill
\resizebox{\columnwidth}{!}{\rotatebox{0}
{\includegraphics{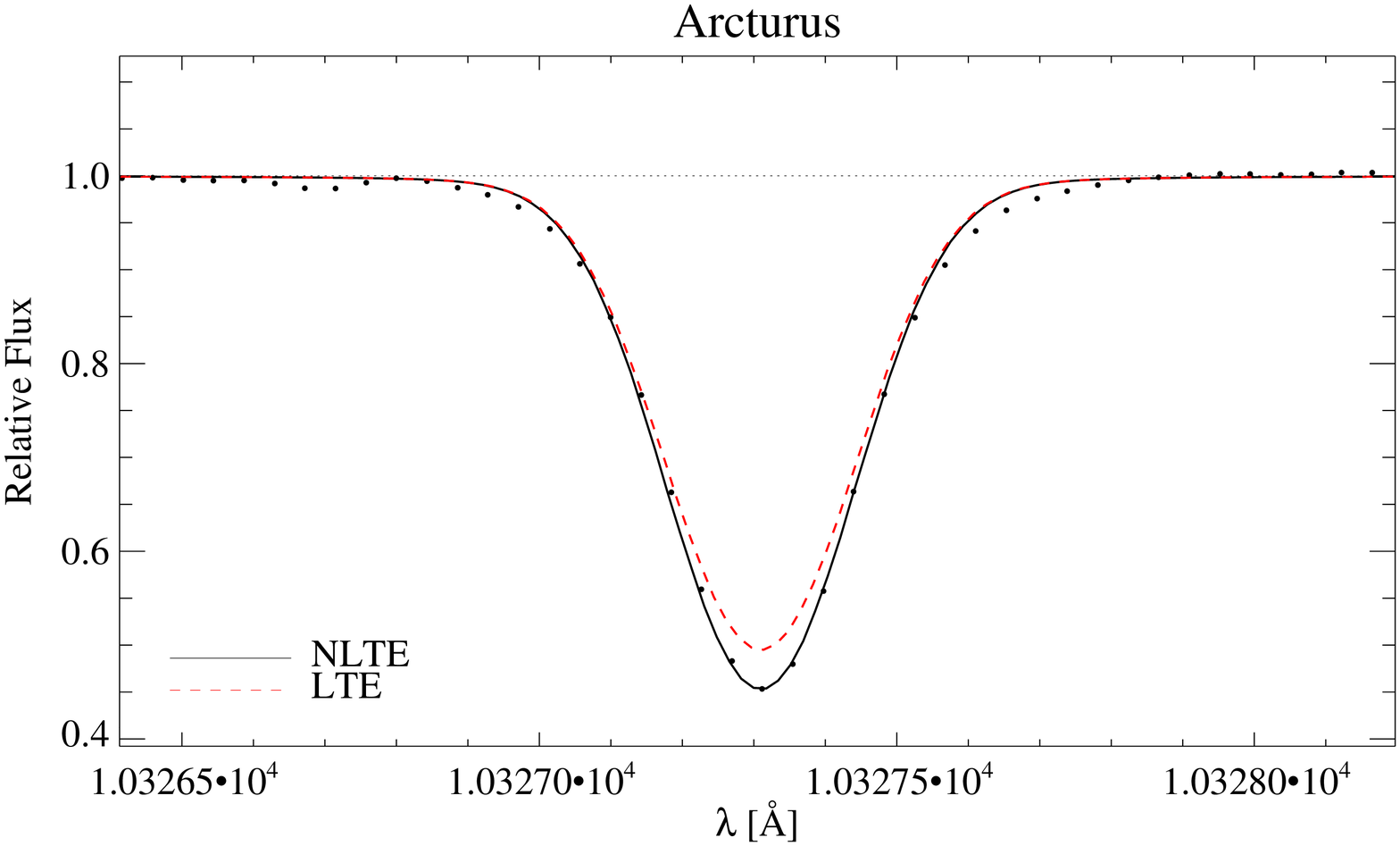}}}}
\hbox{
\resizebox{\columnwidth}{!}{\rotatebox{0}
{\includegraphics{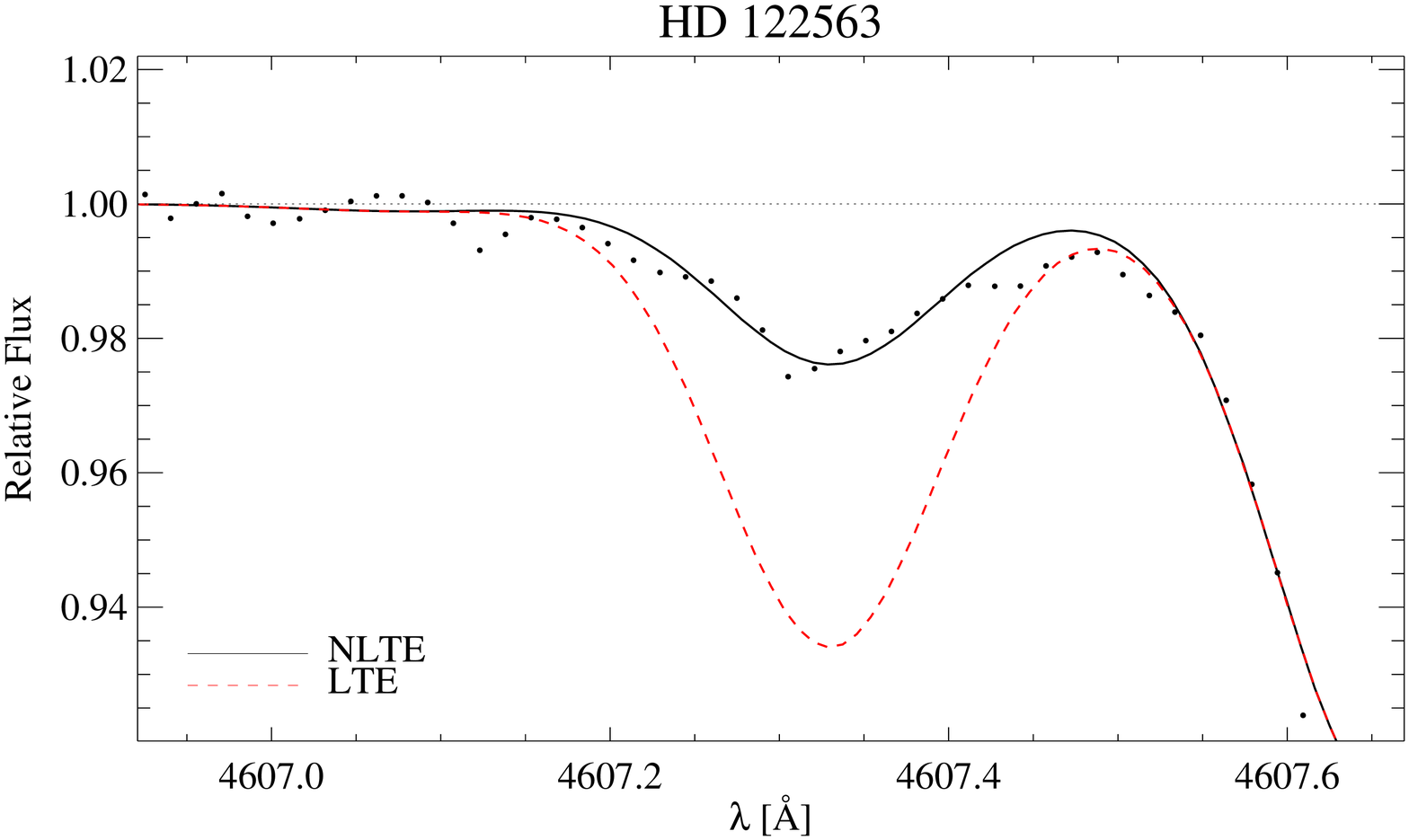}}}\hfill
\resizebox{\columnwidth}{!}{\rotatebox{0}
{\includegraphics{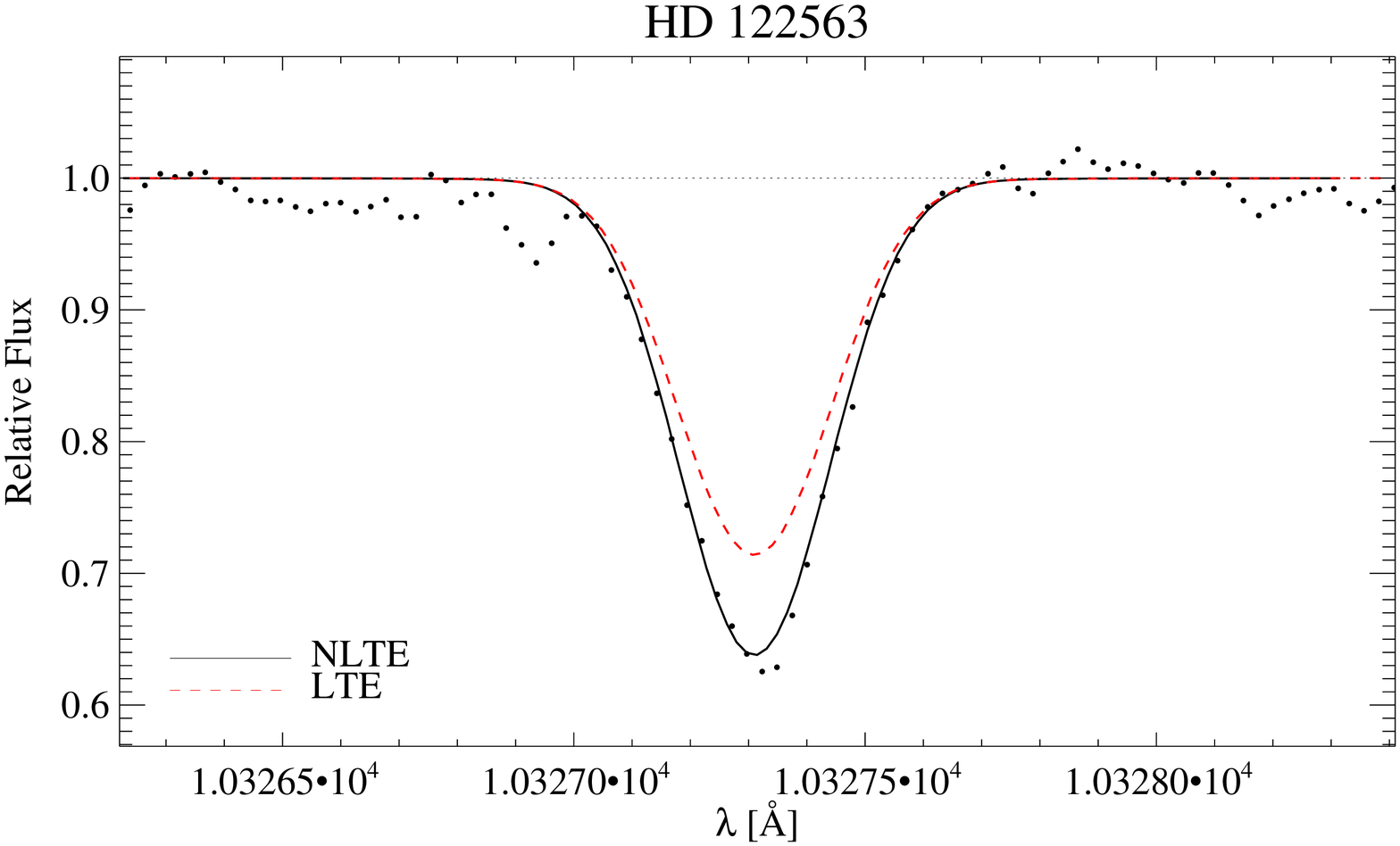}}}}
\caption{NLTE and LTE line profiles of the \ion{Sr}{i} and \ion{Sr}{ii} lines
for selected stars compared with observed stellar spectra.}
\label{profiles}
\end{figure*}

The LTE and NLTE Sr abundances determined for the reference stars are given in
Table \ref{sample} and a line-by-line comparison is shown in Fig.
\ref{abundances}. Few examples of synthetic and observed line profiles are shown
in Fig. \ref{profiles}. Comparison of the [\ion{Sr}{i}/Fe] and [\ion{Sr}{ii}/Fe]
from Table \ref{sample} reveals that the LTE approach predicts a systematic
discrepancy between the lines of two ionization stages, which is also evident
from the Fig. \ref{abundances}. The difference between the \ion{Sr}{i} and the
near-IR \ion{Sr}{ii} lines, which are almost insensitive to damping, is of the
order $0.3$ to $0.5$ dex. This is far beyond the uncertainties in stellar
parameters, which are $\sim 80$ K for $\Teff$, $0.1$ dex for $\log g$, and $0.1$
dex for [Fe/H]. Also, the uncertainties of the transition probabilities are far
smaller than that (Sect. \ref{sec:atdata}).

The NLTE model atom provides, in contrast, realistic ionization balance,
significantly reducing the scatter between the Sr lines of different ionization
stages and excitation energies (Fig. \ref{abundances}). In this respect,
especially important is a good agreement between the NLTE abundances from the
near-IR \ion{Sr}{ii} and visual resonance \ion{Sr}{i} line, which are very
sensitive to NLTE and the effects are of different nature, i.e.,
overionization-stipulated decrease of opacity in the $4607$ line and source
function depletion in the lines of the $1 \mu$m triplet.

\begin{figure}
\begin{center}
\resizebox{\columnwidth}{!}{\includegraphics{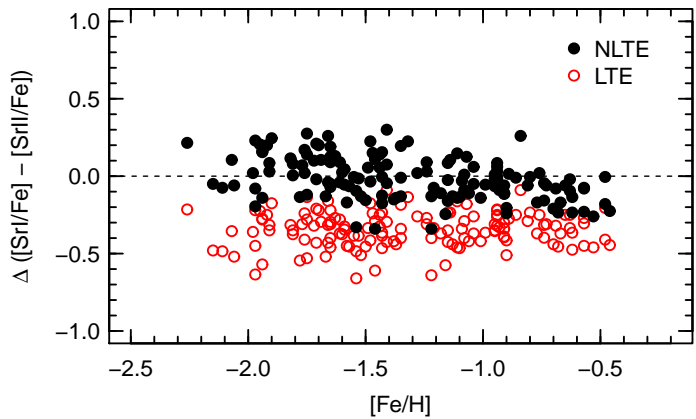}}
\caption{NLTE and LTE abundance differences between \ion{Sr}{i} and \ion{Sr}{ii}
lines for the representative sample of thick-disk and halo stars from Ruchti et
al. (2011).}
\label{ruc}
\end{center}
\end{figure}

The analysis of a larger sample of stars supports the importance of NLTE effects
in the ionization balance of Sr. This is evident from Fig. \ref{ruc}, which
shows the difference between the mean abundances computed using the \ion{Sr}{i}
and \ion{Sr}{ii} lines for a sample of thick-disk and halo RAVE stars with
high-resolution spectroscopic follow-up observations (see Ruchti et al. 2011).
In LTE, the mean offset between the two ionization stages is $\sim 0.4$ dex and
it is alleviated when the NLTE effects are taken into account. A detailed
analysis of the data will be presented in Ruchti et al. (in prep.).

We thus conclude that NLTE must be taken into account in abundance analysis of
resonance \ion{Sr}{i} and subordinate \ion{Sr}{ii} lines in spectra of late-type
stars. Also, at low metallicities, the resonance \ion{Sr}{ii} lines are
sensitive to NLTE effects, and abundances determined in LTE are largely
under-estimated. It is also important to note that the NLTE effects for Sr might
be larger than our estimates, because of interlocking with NLTE-affected lines
of other atoms, as demonstrated by \citet{2006ApJ...641..494S}.

\section{Conclusions}

We investigate statistical equilibrium of Sr in the atmospheres of late-type
stars. The NLTE model atom was constructed using new atomic data, computed with
the R-matrix method, which include levels, transition probabilities,
photoionization and electron-impact excitation cross-sections.

The NLTE effects are significant for the \ion{Sr}{i} resonance line at $4607$
\AA\ and are stipulated by overionization. For the resonance \ion{Sr}{ii}
lines, typically observed in metal-poor FGK stars, our model predicts small
deviations from LTE, although NLTE effects become important at very low
metallicity, [Fe/H] $< -3$, reaching $+0.2$ dex for very metal-poor turnoff
stars. The near-IR \ion{Sr}{ii} triplet shows substantial sensitivity to NLTE,
and the NLTE corrections vary between $+0.1$ and $-0.2$ dex depending on the
model metallicity, temperature, and surface gravity.

The NLTE model atom was applied to the analysis of the Sun and three stars with
well-constrained stellar parameters. In contrast to LTE approach, the NLTE model
recovers ionization and excitation equilibria of Sr for all these reference
stars, thus confirming that the NLTE modeling approach developed in this work
provides a solid base for future abundance determinations of late-type stars. 

The NLTE abundance corrections are provided for the important diagnostic
\ion{Sr}{i} and \ion{Sr}{ii} lines for a grid of model atmospheres in the
following range of stellar parameters $4400 \leq \Teff \leq 6000$ K, $2.2 \leq
\log g \leq 4.6$, $-4 \leq$ [Fe/H] $\leq 0$; for other Sr lines and other models
the abundance corrections can be also computed by request.

\begin{acknowledgements}
Based on observations made with the European Southern Observatory telescopes
(obtained from the ESO/ST-ECF Science Archive Facility). We thank Luca Sbordone
for help with the revision of SIU.
\end{acknowledgements}
\bibliographystyle{aa}

\end{document}